\documentclass[3p,preprint]{elsarticle}

\usepackage{lineno,hyperref}

\journal{Acta Materialia}

\usepackage{mathtools, cuted}  
\usepackage{color, soul} 
\usepackage[colorinlistoftodos, textsize=scriptsize]{todonotes} 
\usepackage{amsmath}
\usepackage{xfrac}  
\usepackage{graphicx}
\usepackage{adjustbox}
\usepackage{gensymb}
\usepackage{makecell}  
\usepackage{array} 
\usepackage{array}  
\usepackage{stackengine}  
\usepackage{multirow}  
\usepackage{resizegather}  
\usepackage{ulem} 
\newcolumntype{L}[1]{>{\raggedright\let\newline\\\arraybackslash\hspace{0pt}}m{#1}}
\newcolumntype{C}[1]{>{\centering\let\newline\\\arraybackslash\hspace{0pt}}m{#1}}
\newcolumntype{R}[1]{>{\raggedleft\let\newline\\\arraybackslash\hspace{0pt}}m{#1}}

\newcommand{\abinitio}{\textit{ab initio} }

\begin{document}

\begin{frontmatter}

\title{Solute diffusion by self-interstitial defects and radiation-induced segregation in ferritic Fe-\textit{X} (\textit{X}=Cr, Cu, Mn, Ni, P, Si) dilute alloys}

\author[KTH, CEA]{Luca Messina\corref{cor1}}
\ead{luca.messina@cea.fr}
\author[CEA]{Thomas Schuler}
\author[CEA]{Maylise Nastar}
\ead{maylise.nastar@cea.fr}
\author[CEA]{Mihai-Cosmin Marinica}
\author[KTH]{P\"ar Olsson}
\cortext[cor1]{Corresponding author}

\address[KTH]{KTH Royal Institute of Technology, Nuclear Engineering, SE--106 91 Stockholm, Sweden}
\address[CEA]{CEA, DEN--Service de Recherches de M\'etallurgie Physique, Universit\'e Paris--Saclay, F--91191 Gif--sur--Yvette, France}

\begin{abstract}
 This work investigates solute transport due to self-interstitial defects and radiation induced segregation tendencies in dilute ferritic alloys, by computing the transport coefficients of each system based on \abinitio calculations of binding energies, migration rates, as well as formation and migration vibrational entropies. The implementation of the self-consistent mean field method in the KineCluE code allows for the calculation of transport coefficients extended to arbitrary interaction ranges, crystal structures, and diffusion mechanisms. In addition, the code gives access to the diffusion and dissociation rates of small solute-defect clusters -- in this case, vacancy- and dumbbell-solute pairs. The results show that the diffusivity of P, Mn, and Cr solute atoms is dominated by the dumbbell mechanism, that of Cu by vacancies, while the two mechanisms might be in competition for Ni and Si, despite the fact that the corresponding mixed dumbbells are not stable. Systematic positive radiation-induced segregation (RIS) at defect sinks is expected for P and Mn solutes due to dumbbell diffusion, and for Si due mainly to vacancy drag. Vacancy drag is also responsible for Cu and Ni enrichment at sinks below 1085 K. The RIS behavior of Cr is the outcome of a fine balance between enrichment due to the dumbbell diffusion mechanism and depletion due to the vacancy one. Therefore, for dilute Cr concentrations global enrichment occurs below 540 K, and depletion above. This threshold temperature grows with solute concentration. The findings are in qualitative agreement with experimental observations of RIS and clustering phenomena, and confirm that solute-defect kinetic coupling plays an important role in the formation of solute clusters in reactor pressure vessel steels and other alloys.
\end{abstract}


\end{frontmatter}


\section{Introduction}
Atomic transport plays a key role in driving the evolution of structural properties of metals and alloys during fabrication, processing, and operation. For instance, solute diffusion determines the final microstructure during phase transformations in heat treatments \cite{clouet_complex_2006, danoix_atom_2006, mao_mechanism_2007}, and the knowledge of the atomic diffusion mechanisms is necessary to adjust the process parameters and obtain the desired properties. Atomic diffusion controls also the microstructure evolution of irradiated materials driven by the permanent excess of point defects (PD). Due to kinetic coupling between PD and atomic fluxes, solutes can migrate to PD sinks under the effect of thermodynamic driving forces acting only on PD. This may cause radiation-induced segregation (RIS) on grain boundaries, dislocations, PD clusters, surfaces, or precipitate-matrix interfaces \cite{nastar_radiation-induced_2012, ardell_radiation-induced_2016, thuinet_multiscale_2018}.  In turn, RIS can catalyze solute-defect clustering and be the precursor of phase precipitation in otherwise undersaturated alloys, if saturation conditions are reached locally. Several studies have shown for instance the formation of solute-enriched clusters (containing Cr, P, Mn, Ni, and Si) in undersaturated FeCr alloys \cite{pareige_behaviour_2015, gomez-ferrer_role_2019} as well as in reactor pressure vessel (RPV) model alloys and steels \cite{nishiyama_effects_2008, lambrecht_positron_2009, meslin_irradiation-induced_2011, meslin_radiation-induced_2013, miller_atom_2013, wells_evolution_2014, styman_post-irradiation_2015}, and even the precipitation of a secondary phase in an ion-irradiated Fe-3\%Ni alloy \cite{belkacemi_radiation-induced_2018} well below the Ni solubility limit. The thermal stability of Mn-Ni-Si precipitates in RPV steels, possibly enhanced by a locally low Fe content and by the presence of irradiation-induced defects \cite{bonny_monte_2014, king_formation_2018}, has been predicted by thermodynamics-informed models and simulations \cite{xiong_thermodynamic_2014, ke_thermodynamic_2017, bonny_impact_2019} and confirmed experimentally \cite{wells_evolution_2014, sprouster_structural_2016, almirall_elevated_2019, almirall_mechanistic_2020}. In addition to the thermodynamic driving forces at play, solute-defect kinetic coupling can contribute to the formation of solute-rich precipitates by increasing the solute concentration at sinks. Moreover, in under-saturated solid solutions with no driving forces for precipitation, it can provide a local driving force whenever RIS leads to local solute concentrations above the solubility limit. Due to this interplay between thermodynamics and kinetics, a radiation-enhanced mechanism for nucleation and growth of solute precipitates may arise. For instance, defect clusters immobilized by small amounts of solutes can turn into nucleation sites where other solutes are progressively accumulated by PD-assisted transport \cite{bonny_monte_2014}. This is confirmed by several observations of toroidal-shaped and planar solute clusters in RPV steels \cite{miller_embrittlement_2000, meslin_radiation-induced_2013, altstadt_fp7_2014, marquis_atom_2015}, and has been recently rationalized in an advanced model able to reproduce the microstructure evolution of various RPV steel types \cite{castin_dominating_2020}. The "pinning" effect on clusters of self-interstitial atoms (SIA) has been inferred from the interpretation of experimental results for Mn, P, and Ni \cite{meslin_radiation-induced_2013, pareige_behaviour_2015, belkacemi_radiation-induced_2018, gomez-ferrer_role_2019}, and by modeling for Cr, P, and Ni \cite{terentyev_correlation_2005, meslin_theoretical_2007, osetsky_role_2015}. Recent \abinitio calculations of solute-loop interactions suggest a similar pinning tendency for Mn, Cu, and Si \cite{domain_solute_2018}.	

However, the capability of PDs (vacancies and self-interstitials) to carry solute atoms to the nucleation sites is yet not fully characterized. Nowadays, precise analytical models based on the Self-Consistent Mean Field (SCMF) theory \cite{nastar_mean_2005} or the Green-function approach \cite{trinkle_automatic_2017}, in combination with \abinitio calculations of defect jump rates, allow for a highly accurate analysis of the intrinsic atomic-transport properties by calculation of the transport (Onsager) matrix \cite{allnatt_atomic_2003}. By the latter methods, it has been proven that solute drag by vacancies is a widespread phenomenon arising below a given temperature threshold in body-centered cubic (bcc) \cite{garnier_solute_2013, messina_systematic_2016}, face-centered cubic (fcc) \cite{garnier_calculation_2014, garnier_diffusion_2014, garnier_quantitative_2014}, and hexagonal close-packed (hcp) metals \cite{agarwal_exact_2017}, provided that the vacancy-solute interaction is sufficiently strong. In dilute ferritic alloys, the threshold temperature has been systematically determined for all transition-metal impurities \cite{messina_exact_2014, messina_systematic_2016}. This threshold is near or above 1000 K for Cu, Mn, Ni, P, and Si, whereas it lies near 300 K for Cr. Therefore, vacancies are capable of transporting all solutes to sinks, with the exception of Cr for which depletion at sinks is expected.

On the other hand, the transport efficiency of self-interstitial atoms (SIA) has been only superficially investigated. Speculations based on \abinitio evaluations of the stability of mixed dumbbells (MD) \cite{domain_diffusion_2005, vincent_ab_2006, olsson_ab_2010}, and the interpretation of resistivity-recovery (RR) experiments \cite{maury_study_1987, maury_interstitial_1990, abe_interaction_1999} lead to the assumption that transport of Cr, Mn, and P should occur, since the corresponding MDs are stable, whereas for the opposite reason it should not occur for Ni, Cu, and Si. As a matter of fact, for Ni and Si conflicting assumptions and interpretations are found in the literature \cite{maury_interstitial_1990, abe_interaction_1999, pareige_behaviour_2015}. Sometimes, even more simplified assumptions are made based only on solute size, namely that enrichment is expected for undersized solutes, and depletion for oversized ones \cite{lu_irradiation-induced_2008}. Only Cr and Ni solutes have been addressed in more detail with full calculations of transport coefficients. Choudhury \textit{et al.} \cite{choudhury_ab-initio_2011} found that no transport of Ni by SIAs should occur. However, the drawn conclusions on the RIS behavior disregarded the effects of flux coupling and predicted Ni depletion, in disagreement with the experimental evidence \cite{belkacemi_radiation-induced_2018}. The FeCr alloy has been the object of a series of studies based on rate-theory \cite{wharry_mechanism_2014}, atomistic kinetic Monte Carlo (AKMC) \cite{senninger_modeling_2016}, and phase-field \cite{piochaud_atomic-based_2016, thuinet_multiscale_2018} models. They showed that SIA transport plays a crucial role and yields Cr enrichment at sinks in opposition to the vacancy-induced depletion. The balance of the two effects provides an explanation for the experimentally-observed \cite{wharry_systematic_2013} existence of a switchover temperature between enrichment and depletion, and has been shown to be extremely sensitive to microstructure features (e.g., composition, strain fields, or sink density) \cite{senninger_modeling_2016, martinez_role_2018}. On the contrary, no studies have yet fully uncovered the reasons for the systematic enrichment tendencies observed for the other solutes (Cu, Mn, Ni, P, Si) on grain boundaries, dislocations lines, and loops \cite{wharry_systematic_2013, meslin_radiation-induced_2013, miller_atom_2013, pareige_behaviour_2015}. It is unclear whether this is due to vacancy drag, SIA transport, or a combination of both.

The objective of this study is to perform an accurate analysis of SIA-driven solute transport in a set of dilute Fe-$X$ alloys ($X$=Cr, Cu, Mn, Ni, P, Si), and provide a comprehensive overview of the expected intrinsic RIS behavior of said chemical species. In addition, this work establishes a general modeling framework aimed at accurate RIS predictions by vacancies and interstitials in dilute alloys, providing in this way a sound knowledge of the kinetic properties necessary to the interpretation of diffusion-driven microstructure phenomena. This is achieved by means of the recently published KineCluE code \cite{schuler_kineclue:_2020} that implements the SCMF method in its cluster-expansion form \cite{schuler_transport_2016}. For SIA diffusion, KineCluE allows for a considerable advancement with respect to the available analytical models \cite{barbe_phenomenological_2006, barbe_phenomenological_2007} developed in the SCMF framework and featuring limited solute-SIA interaction ranges. KineCluE generalizes this framework to arbitrarily long ranges, solute concentration effects, and complex SIA migration mechanisms in any periodic crystal. In the next sections, after a brief overview over the cluster-expansion framework, KineCluE is applied to SIA-solute migration in the aforementioned bcc ferritic alloys, based on accurate \abinitio calculations of binding energies and migration rates. The Onsager matrices are then combined with the vacancy-related transport coefficients \cite{messina_exact_2014} and used to analyze flux coupling and RIS tendencies, in comparison with the current experimental and modeling knowledge.

\section{Methodology for the calculation of transport coefficients}

\subsection{Cluster expansion of transport coefficients}
In the framework of linear thermodynamics of irreversible processes, transport coefficients relate atomic fluxes to the underlying thermodynamic forces, e.g., chemical potential gradients (CPG) \cite{allnatt_atomic_2003}. The symmetric Onsager matrix allows for the analysis of flux coupling between different species, for instance PD-induced solute transport. The components $L_{ij}$ are defined as:
\begin{equation}
J_i = - \sum_j L_{ij} \frac{\nabla \mu_j}{k_\mathrm{B}T} \; ,
\label{eq:flux_equation}
\end{equation}
where $J_i$ is the flux of species $i$ and $\nabla \mu_j$ the CPG of species $j$. In binary alloys, the independent coefficients needed for each PD type are $L_\mathrm{\delta \delta}$, $L_{\delta \mathrm{B}}$, and $L_\mathrm{BB}$, where $\delta = (\mathrm{V,I})$ marks vacancies or interstitials, and $\mathrm{B}$ the solute species. The solvent (A) coefficients $L_\mathrm{AA}$,  $L_\mathrm{AB}$, and  $L_{\delta \mathrm{A}}$ can be inferred from the former (cf. Table \ref{tab:formula_summary}).

The transport coefficients are computed here within the SCMF theory \cite{nastar_mean_2005}. The latter has been applied multiple times to vacancy diffusion \cite{garnier_solute_2013, garnier_quantitative_2014, messina_exact_2014, messina_systematic_2016, schuler_transport_2016, claisse_transport_2016}, but the only model available for SIA-assisted solute transport is that by Barbe \textit{et al.} \cite{barbe_phenomenological_2006, barbe_phenomenological_2007, barbe_split_2007, barbe_split_2007-1}. The latter is here substantially improved with the KineCluE code \cite{schuler_kineclue:_2020}, which implements a recent development of SCMF where the Onsager matrix is expanded in terms of cluster contributions $L_{ij}^{(c)}$ \cite{schuler_transport_2016}: 
\begin{equation}
L_{ij} = \mathcal{C} \sum_c f_c L_{ij}^{(c)} \; .
\label{eq:Lij_cluster_development}
\end{equation}
$\mathcal{C}$ is the total concentration, while $f_c$ is the concentration fraction of cluster $c$ proportional to the cluster partition function (cf. \ref{sec:appendix_computation_total_lij}). In the dilute limit, the matrix is split into a "monomer" (an isolated defect) and a (solute-defect) "pair" contribution, while immobile species such as substitutional solutes do not contribute to the total transport matrix. Cluster transport coefficients and partition functions are computed with KineCluE and then combined with a concentration model to obtain the cluster fractions appearing in Eq. \ref{eq:Lij_cluster_development}. In the case of solute transport by vacancies and interstitials in a dilute binary alloy, the total Onsager matrix reads \cite{schuler_design_2017}:

	\begin{equation}
	\begin{bmatrix}
	L_\mathrm{VV} &  L_\mathrm{VI}  & L_\mathrm{VB} \\ L_\mathrm{VI}  &  L_\mathrm{II}  &  L_\mathrm{IB}  \\  L_\mathrm{VB}  & L_\mathrm{IB}  &  L_\mathrm{BB}
	\end{bmatrix}
	= \mathcal{C} \left( f_\mathrm{V}
	\begin{bmatrix}
		L_\mathrm{VV}^\mathrm{(V)} &  0  &  0  \\  0  &  0  &  0  \\  0  &  0  & 0
	\end{bmatrix}
	+ f_\mathrm{I}
	\begin{bmatrix}
		0 & 0 & 0 \\ 0  &  L_\mathrm{II}^\mathrm{(I)}  &  0  \\  0 & 0 & 0
	\end{bmatrix}
	+ f_\mathrm{VB}
	\begin{bmatrix}
		L_\mathrm{VV}^\mathrm{(VB)} & 0  & L_\mathrm{VB}^\mathrm{(VB)} \\ 0  & 0  & 0  \\  L_\mathrm{VB}^\mathrm{(VB)}  & 0  &  L_\mathrm{BB}^\mathrm{(VB)}
	\end{bmatrix}
	+ f_\mathrm{IB}
	\begin{bmatrix}
		0 & 0 & 0 \\ 0  &  L_\mathrm{II}^\mathrm{(IB)}  &  L_\mathrm{IB}^\mathrm{(IB)}  \\ 0  & L_\mathrm{IB}^\mathrm{(IB)}  &  L_\mathrm{BB}^\mathrm{(IB)}
	\end{bmatrix} \right) \; .
	\label{eq:cluster_expansion}
	\end{equation}
The vacancy and interstitial contributions to the $L_\mathrm{BB}$ coefficient, respectively $L_\mathrm{BB}(\mathrm{V})=\mathcal{C}f_\mathrm{VB}L_\mathrm{BB}^\mathrm{(VB)}$ and $L_\mathrm{BB}(\mathrm{I})=\mathcal{C}f_\mathrm{IB}L_\mathrm{BB}^\mathrm{(IB)}$, are computed separately. The $L_\mathrm{VI}$ coefficient is set to zero, which means that any kinetic coupling between vacancies and interstitials is neglected. The concentration model and the procedure to compute the total Onsager coefficients are explained in detail in \ref{sec:appendix_computation_total_lij}. Note that in case detailed balance is not satisfied, e.g., as an effect of ballistic atomic relocation under irradiation at low temperature \cite{huang_atomic_2019}, the cluster transport matrices are not symmetric.

KineCluE extends the reach of SCMF to an arbitrarily long kinetic radius $r_\mathrm{kin}$, which defines the maximum extent of the kinetic trajectories included in the calculation. The transport coefficients converge with increasing $r_\mathrm{kin}$ to an asymptotic value \cite{schuler_kineclue:_2020}. After a specific convergence study on SIA-solute migration, the coefficients were found to be well converged at $r_\mathrm{kin} = 4\,a_0$ (a$_0$ is the lattice parameter), and to reach an acceptable accuracy already at $r_\mathrm{kin} = 2\,a_0$, whereas they are not converged with a kinetic radius as short as the first nearest-neighbor (nn) distance (corresponding to Barbe's model). The kinetic radius is thus set here to $4\,a_0$, unless otherwise specified. 

For each cluster, the KineCluE calculation proceeds in three steps:
\begin{enumerate}
	\item   A "symbolic run" that depends only on the crystal geometry and the defect migration mechanisms. This yields the $L_{ij}$ semi-analytical expressions, as well as the list of symmetry-unique configurations and jumps required to perform the numerical calculations, within the interaction radius of choice (cf. Sec. \ref{sec:configurations_jumps}).
	\item  The calculation of the binding and migration energy for each item in the lists, here obtained with \abinitio calculations (cf. Sec. \ref{sec:ab_initio_method}).
	\item  A "numeric run" to compute the cluster transport coefficients and the partition function in a given temperature range (cf. Sec. \ref{sec:solute_transport}).
\end{enumerate}

\subsection{Equilibrium configurations and jump frequencies}
\label{sec:configurations_jumps} 

\begin{figure}[!htb]
	\centering
	\resizebox{0.6\textwidth}{!}{
		\includegraphics{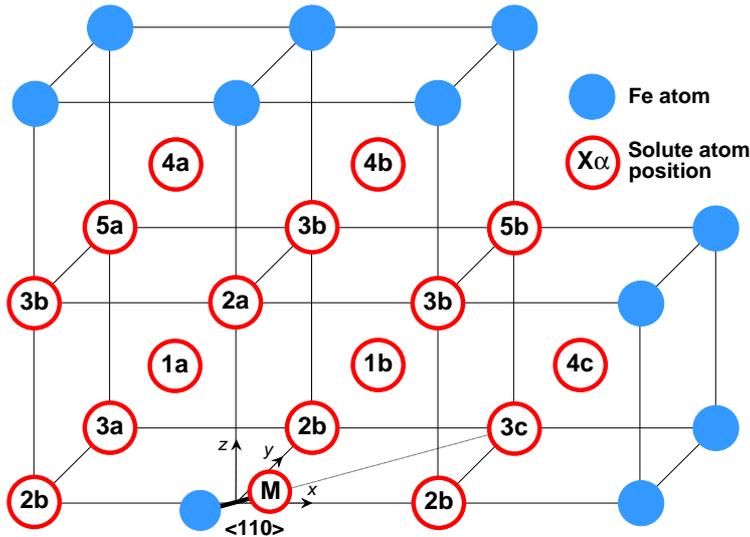}}
	\caption{\label{fig:configurations}Nomenclature of solute-dumbbell equilibrium configurations. The red circles mark the solute position relative to the dumbbell defect: $x$ is the solute nearest-neighbor shell with respect to the dumbbell, and $\alpha$ the symmetry class within the same shell (see Table \ref{tab:dumbbell_coords}). The colored atoms are located outside the interaction shells. 'M' marks the mixed-dumbbell configuration. }
\end{figure}

\begin{figure}[!htb]
	\centering
	\resizebox{\textwidth}{!}{
		\includegraphics{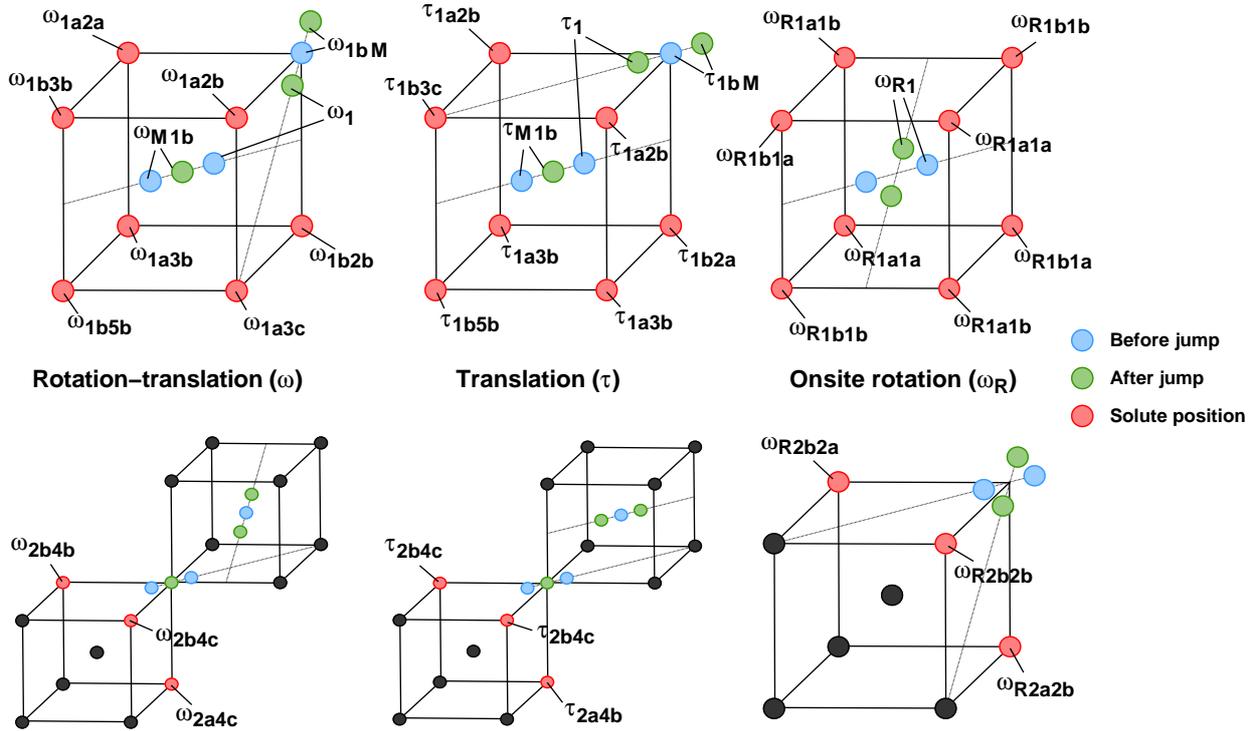}}
	\caption{\label{fig:jump_nomenc} Representation of all dumbbell jumps (rotation-translations $\omega$, translations $\tau$, and onsite rotations $\omega_\mathrm{R}$) connecting the solute-dumbbell configurations in Fig. \ref{fig:configurations}, i.e., for an interaction range $r_\mathrm{th}$ extending to the 5nn shell.}
\end{figure}

In bcc iron, the most stable SIA configuration is the $\langle 110 \rangle$ dumbbell \cite{fu_stability_2004, domain_diffusion_2005}, with three distinct jump mechanisms \cite{bocquet_solute-and-dumbbell_1991, barbe_phenomenological_2006}: 60\degree{} rotation-translation jumps (RT) to four target sites out of the eight 1nn, pure-translation jumps (T) to the same target sites, and 60\degree{} onsite rotations (R). The target sites, labelled '1b' in Fig. \ref{fig:configurations}, are those in a so-called "compressed" position. In addition, it is possible to have a 2nn 90\degree{} jump \cite{fu_stability_2004}. For a solute-dumbbell pair, it is necessary to define two additional jump mechanisms, namely the MD migration and dissociation. The former Barbe's model limited to 1nn interactions is here generalized by extending the kinetic radius to $4 \, a_0 $, which yields a list of 94 symmetry-unique configurations and 260 symmetry-unique RT jump frequencies, compared to the 3 and 8 featured in Barbe's model. This allows for a much better precision in the transport coefficient calculation, thanks to the inclusion of wider solute-dumbbell correlated trajectories. The computational load can be then reduced by setting a smaller thermodynamic radius, which depends on the extent of the thermodynamic interaction between dumbbell and solute. Consistently with the \abinitio binding energies (cf. Table \ref{tab:DFT_results}), it is chosen to set $r_\mathrm{th}$ to the 5nn distance. This yields a subset of 13 configurations and 13 RT jump frequencies that are computed by precise \abinitio calculations, and are shown in Figs. \ref{fig:configurations} and \ref{fig:jump_nomenc}. 

In Fig. \ref{fig:configurations}, sites in the same nn shell are categorized depending on the distance from each atom of the dumbbell. Each site is thus labeled '$X\alpha$', where $X$ stands for the nn shell, and $\alpha$ for the specific symmetry class within that shell (with the exception of the MD configuration, marked as 'M'). For each symmetry class, sample atomic coordinates are provided in \ref{sec:appendix_coordinates}. In Fig. \ref{fig:jump_nomenc}, the jumps connecting these configurations are marked with $\omega$, $\tau$, and $\omega_\mathrm{R}$, for RT, T, and R jumps, respectively. The MD jumps are marked with '1', while the jumps of the Fe-Fe dumbbell are either labeled with '$0$', when far away from solute atoms, or according to the initial ($X\alpha$) and final ($Y \beta$) configuration when nearby solutes. Refer to Table \ref{tab:allowed_jumps} for a summary.

The jumps shown in Fig. \ref{fig:jump_nomenc} are modeled according to transition-state theory (TST) as thermally-activated processes with frequency \cite{allnatt_atomic_2003}:
\begin{equation}
\omega_{ij} = \nu_{ij} \exp \left( - \frac{E^\mathrm{mig}_{ij}}{k_B T}  \right) \; ,
\label{eq:jump_freq}
\end{equation}
where $\nu$ is the attempt frequency and $E^\mathrm{mig}$ the migration energy. Beyond $r_\mathrm{th}$, the binding energies are set to zero, and the jump frequencies are obtained with a classic kinetically-resolved activation (KRA) approach \cite{van_der_ven_first_2005}, which was proven sufficiently accurate for far-away jumps \cite{messina_exact_2014}: 
\begin{equation}
E^\mathrm{mig}_{ij} = E^\mathrm{mig}_0 + \frac{E^\mathrm{b}_i - E^\mathrm{b}_j}{2} \; .
\label{eq:kra}
\end{equation}
$E^\mathrm{b}_i$ and $E^\mathrm{b}_j$ are the binding energy (positive when attractive) of the initial and final state, and $E^\mathrm{mig}_0$ is the migration barrier of the $\omega_0$ RT jump. This approach with a reduced $r_\mathrm{th}$ allows for a better accuracy with the same amount of \abinitio calculations. Furthermore, detailed-balance requirements are automatically fulfilled. 

\subsection{Ab initio methodology}
\label{sec:ab_initio_method}

The binding and migration energies are calculated in the framework of Density Functional Theory (DFT) with the Vienna \textit{ab initio} Simulation Package ({\sc vasp}) \cite{kresse_ab_1993, kresse_ab_1994, kresse_norm-conserving_1994}. The {\sc vasp} full-core pseudopotentials developed within the projector augmented wave (PAW) method \cite{blochl_projector_1994, kresse_ultrasoft_1999} are employed for all elements. The Perdew-Burke-Ernzerhof (PBE) parameterization \cite{perdew_generalized_1996} of the generalized gradient approximation (GGA) is used to describe the exchange-correlation function. Calculations are spin-polarized, and make use of the Vosko-Wilk-Nusair (VWN) interpolation scheme of the correlation potential. Particular attention has been paid to achieve the correct antiferromagnetic state of Mn and avoid local minima \cite{schneider_local_2018}, by providing a precise initial guess of the magnetic moment on Fe ($2.23\,\mu_\mathrm{B}$) and Mn ($-2.24\,\mu_\mathrm{B}$) atoms \cite{domain_solute_2018}, and by applying linear mixing in the starting guess of the charge dielectric function \cite{messina_systematic_2016}. 

Ionic relaxations are performed in a 251-atom supercell, whose volume and shape are maintained to that of a bcc iron cell. Following previous convergence studies \cite{olsson_ab_2007}, the equilibrium lattice parameter is found to be $a_0=2.831\,\text{\AA}$. The Brillouin zone is sampled with a Monkhorst-Pack $3\times 3\times 3$ \textit{k}-point grid, and a plane-wave cut-off of 350 eV is used. More accurate parameters ($4\times 4\times 4$ k-points and 400 eV cut-off) are used to refine the calculations with Cr atoms, because the energy differences among distinct configurations are small and the results are thus more sensitive to the input parameters. 

The dumbbell formation energy is:
\begin{equation}
	E^\mathrm{f} = E[N+1]-\frac{N+1}{N} E[N] \; ,
\end{equation}  
where $E[N+1]$ and $E[N]$ are the energies of the supercell with and without the dumbbell, respectively. The binding energies of the solute-dumbbell configurations shown in Fig. \ref{fig:configurations} are evaluated as:
\begin{equation}
E_{X\alpha}^\mathrm{b} = - E[\mathrm{I,B}] + E[\mathrm{I}] + E[\mathrm{B}] -  E_\mathrm{ref} \; ,
\end{equation}
where the terms on the right-hand side are the energy of the supercell containing one dumbbell and one solute, one dumbbell, one solute, and no defect, respectively. In this convention, positive signs stand for attractive interactions.

The migration barriers (Eq. \ref{eq:jump_freq}) are obtained by means of nudged-elastic band (NEB) calculations \cite{mills_reversible_1995, jonsson_nudged_1998}, implemented with the climbing-image algorithm \cite{henkelman_climbing_2000} and three intermediate images. In order to reduce the computational load, calculations are mainly limited to the RT mechanism, as the latter is characterized by the lowest migration barriers. The accuracy on each migration barrier is estimated in less than 5 meV. 

The attempt frequencies of the Fe-Fe dumbbell ($\omega_0$) and the MDs RT jumps ($\omega_1$) are computed by means of DFT frozen-phonon calculations \cite{vineyard_frequency_1957, lucas_vibrational_2009, messina_systematic_2016} in 128-atom cells, with a cut-off energy of 400 eV. All other jumps shown in Fig. \ref{fig:jump_nomenc} are assigned the attempt frequency of $\omega_0$. The vibrational frequencies of each dumbbell in the equilibrium and saddle-point positions $\nu_k^\mathrm{E}$ and $\nu_k^\mathrm{S}$ are obtained in two steps: first, by relaxing the configuration with a force convergence criterion of 0.01 eV/\AA{} on each atom, and restraining the supercell volume to the value stemming from the equilibrium lattice parameter $a_0$ mentioned above. Secondly, by applying four displacements of $\pm 0.010$ and $\pm 0.020\,\text{\AA}$ on each atom and diagonalizing the Hessian matrix. These values ensure an accurate numerical estimation of the Hessian and its phonon spectrum \cite{marinica_energy_2011}. The attempt frequency is given by \cite{vineyard_frequency_1957}:
\begin{equation}
\nu = \frac {\prod_{k=1}^{3N_\mathrm{E}-3}\nu_k^\mathrm{E}}{\prod_{k=1}^{3N_\mathrm{E}-4}\nu_k^\mathrm{S}}
\end{equation}
where $N_\mathrm{E}$ is the total number of atoms in the supercell. In addition, the dumbbell formation entropy in pure Fe can be obtained as \cite{mishin_calculation_2001}:
\begin{equation}
\frac{S^\mathrm{f}}{k_\mathrm{B}} = - \left[ \ln \left( \prod_{k=1}^{3N_\mathrm{E}-3} \nu_k^\mathrm{E} \right) - \frac{N_\mathrm{E}}{N_\mathrm{U}} \ln \left( \prod_{k=1}^{3N_\mathrm{U}-3} \nu_k^\mathrm{U} \right) \right] \; ,
\end{equation}
where $\nu_k^\mathrm{U}$ and $N_\mathrm{U}$ are the eigenfrequencies and the number of atoms in the undefected supercell, respectively.

\subsection{Vacancy transport coefficients}
The vacancy coefficients were already calculated within the SCMF theory in a previous work \cite{messina_exact_2014}, but they are recomputed here with a longer kinetic radius ($r_\mathrm{th} = 4\,\mathrm{a}_0$) and the cluster-expansion approach (Eq. \ref{eq:cluster_expansion}), for the sake of consistency with the SIA calculations. The thermodynamic radius is set to the 5nn distance ($\sqrt{3}\,\mathrm{a}_0$), yielding 12 jump frequencies that correspond exactly to the ones depicted in Fig. 1 of reference \cite{messina_exact_2014}. The DFT calculations are repeated with the same \textit{k}-point grid ($3\times 3\times 3$) but a higher cut-off energy (400 eV in place of 300 eV) and a larger simulation cell (249 atoms instead of 127). No substantial differences are found, with the exception of larger $\omega_{15}$ barriers in FeP and FeSi (0.85 and 0.82 eV instead of 0.72 and 0.71 eV, respectively). Finally, jumps for non-compact configurations (i.e., from $\sqrt{3}$ to $4\,\mathrm{a}_0$) are obtained with the KRA approach (Eq. \ref{eq:kra}). To this purpose, the solute-vacancy binding energies are set to their \abinitio value up to the 10nn ($2.6\,a_0$), and to zero beyond this distance.

\section{DFT results}
\label{sec:dft_results}

\subsection{Dumbbell properties in pure iron}

The dumbbell properties in pure Fe are summarized in Table \ref{tab:DFT_pure}. At 0 K, the $\langle 110\rangle$ orientation, with a formation energy of 4.08 eV, is more favorable than the $\langle 111\rangle$ orientation by a margin of 0.75 eV, in agreement with previous studies performed with other DFT codes or functionals \cite{fu_stability_2004, domain_diffusion_2005}. The $\langle 110 \rangle$ formation entropy (0.05 k\textsubscript{B}) is much lower than that of the Ackland--Mendelev potential (\cite{ackland_development_2004, marinica_orientation_2007}), but closer to Lucas and Sch\"aublin's DFT calculations based on the same functionals and a lower cut-off energy \cite{lucas_vibrational_2009}. According to the latter study, the large formation-entropy difference between $\langle 110\rangle$ (0.05 k\textsubscript{B}) and $\langle 111 \rangle$ dumbbells (4.17 k\textsubscript{B}) sensibly reduces the formation free-energy gap at finite temperatures. 

The migration barriers of the different jump types are in good agreement with previous calculations, confirming that rotation-translation is the most probable mechanism. For a $\langle 110\rangle$--$\langle 011\rangle$ onsite rotation, the saddle-point orientation is $\langle 1h1 \rangle$ ($h\approx 2$). In the RT jump, the hopping atom forms a symmetric double dumbbell with the initial and final atom with orientations $\langle kj1 \rangle$ and $\langle 1jk \rangle$  ($j=3.3$ and $k=5$, approximately). The 2nn jump has a slightly lower barrier (0.46 eV) than previous computations \cite{fu_stability_2004} and is the second most probable jump.  

Finally, the RT attempt frequency of 4.44 THz is close to the Debye frequency in Fe (6 THz) \cite{kittel_introduction_2004} and lower than that of a vacancy jump (10.8 THz) \cite{messina_systematic_2016}. The resulting dumbbell diffusion prefactor $D_0=a_0^2\exp(S^\mathrm{f}/\mathrm{k}_\mathrm{B})\nu_0 = 3.74\cdot10^{-7}$ m\textsuperscript{2}/s is about an order of magnitude lower than the results of molecular-dynamics simulations based on the Ackland--Mendelev potential \cite{anento_atomistic_2010}. The reason is twofold. Firstly, since the latter simulations are performed at much higher temperatures, they account for anharmonic effects. Secondly, even within the harmonic approximation the force field based on the Ackland--Mendelev potential provides for higher vacancy-diffusion prefactors than DFT \cite{athenes_estimating_2012}. Both topics are beyond the scopes of this paper and will be the object of future studies.


\begin{table*}[!htb]\scriptsize
	\caption{Summary of the DFT dumbbell properties in pure Fe, referring to the $\langle 110 \rangle$ orientation unless otherwise specified. }
	\renewcommand{\arraystretch}{1.5}	
	\centering
	\begin{adjustbox}{width=\textwidth}
		\begin{tabular}{L{5cm}L{1.5cm}L{2.5cm}L{7cm}}
			\hline
			\hline
			\textbf{Quantity}																&																			&	\textbf{This work}				&	\textbf{Previous works}		\\
			Lattice parameter																&	$a_0$															&	2.831 \AA							&				\\
			Dumbbell formation entropy												&	$S^\mathrm{f}_{\langle 110 \rangle}$	&	0.050 k\textsubscript{B}		&		0.24 k\textsubscript{B}\textsuperscript{\cite{lucas_vibrational_2009}}, 1.41\textsuperscript{\cite{marinica_orientation_2007}}, 2.84\textsuperscript{\cite{marinica_orientation_2007}}		\\	
			Dumbbell formation enthalpy												&	$E^\mathrm{f}_{\langle 110 \rangle}$	&	4.082 eV									&		3.64\textsuperscript{\cite{fu_stability_2004}}, 3.94\textsuperscript{\cite{domain_ab_2001}}		\\	
			$\langle 111 \rangle $-dumbbell formation enthalpy		&	$E^\mathrm{f}_{\langle 111 \rangle}$	&	4.832 eV									&		4.34\textsuperscript{\cite{fu_stability_2004}}, 4.66 \textsuperscript{\cite{domain_ab_2001}}		\\	
			Rotation-translation energy barrier									&	$\omega_0$												&	0.335 eV									&	0.32\textsuperscript{\cite{takaki_resistivity_1983}}, 0.37\textsuperscript{\cite{vincent_ab_2006}}, 0.34\textsuperscript{\cite{meslin_theoretical_2007}}, 0.34\textsuperscript{\cite{olsson_ab_2007}}, 0.35\textsuperscript{\cite{choudhury_ab-initio_2011}}				\\
			2nn rotation-translation energy barrier							&	$\omega_{0(\mathrm{2nn})}$					&	0.459 eV									&	0.50\textsuperscript{\cite{fu_stability_2004}}	\\
			Translation energy barrier													&	$\tau_0$													&	0.785 eV									&	0.78\textsuperscript{\cite{olsson_ab_2007}}, 0.84\textsuperscript{\cite{choudhury_ab-initio_2011}}, 0.80\textsuperscript{\cite{vincent_ab_2006}}, 0.78\textsuperscript{\cite{meslin_theoretical_2007}}				\\	
			Onsite rotation energy barrier											&	$\omega_\mathrm{R0}$							&	0.611 eV									&	0.63\textsuperscript{\cite{vincent_ab_2006}}				\\
			Rotation-translation attempt frequency						&	$\nu_0$														&	4.445 THz 								&					\\
			Dumbbell diffusion prefactor	&	$D_0$		&		$3.745\cdot 10^{-7}$ m\textsuperscript{2}/s	&	$2.268\cdot 10^{-6}$ m\textsuperscript{2}/s\textsuperscript{\cite{anento_atomistic_2010}} \\
			\hline
			\hline
		\end{tabular}
	\end{adjustbox}
	\label{tab:DFT_pure}
\end{table*}

\subsection{Solute-dumbbell interactions and jump frequencies}

Table \ref{tab:DFT_results} reports the relevant binding energies and migration barriers of dumbbell-solute pairs, following the nomenclature of Figs. \ref{fig:configurations} and \ref{fig:jump_nomenc}. In addition to RT jumps, the table lists the results obtained by previous studies \cite{vincent_ab_2006, meslin_theoretical_2007, olsson_ab_2010} for 2nn jumps leading to a MD displacement ($\omega_{1,\mathrm{2nn}}$) or dissociation ($\omega_{M\mathrm{2b}}$). The complete set of DFT results, including those related to vacancy diffusion, is available in the associated supplementary database \cite{database}. In a few cases, no barrier between the initial and final state is found, namely for athermal transitions where the end-point energy difference is large. 
In FeP, since configuration '2b' is unstable and relaxes spontaneously to a MD, the related barriers have not been computed. 

The binding energies are in good agreement with previous DFT calculations that were limited to 1nn sites \cite{domain_diffusion_2005, olsson_ab_2010, vincent_ab_2006, meslin_theoretical_2007}. They fade already at the 2nn shell, with a few exceptions. The most important one is the '5b' configuration in FeP, with a non negligible binding (+0.21 eV) due to the fact that the in-between atom (along the $\langle 111 \rangle$ direction) is pushed towards P and approaches a MD conformation. As the MD can migrate in a fully 3D path without dissociating, solute diffusion can take place even in the absence of a strong 1nn binding, as opposed to vacancy drag where 1nn binding is a necessary condition, and 2nn binding a strong additional enhancement factor \cite{messina_systematic_2016}.  

\begin{figure*}[!htb]
	\resizebox{\textwidth}{!}{
		\includegraphics{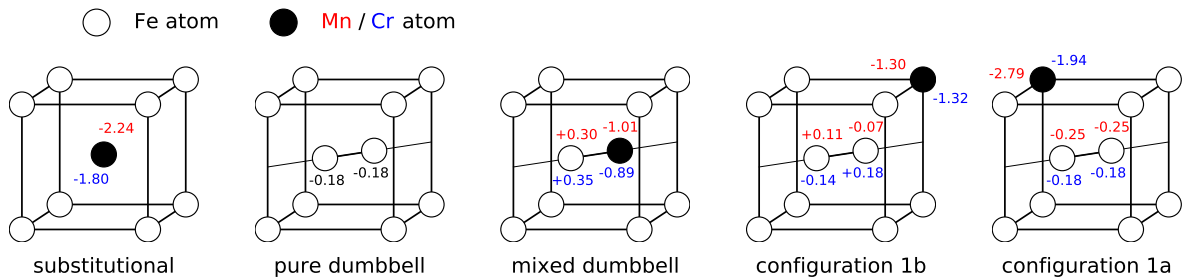}}
	\caption{\label{fig:afm_moments} DFT local magnetic moments (in $\mu_\mathrm{B}$) in different dumbbell-solute configurations, in the presence of Mn (red values) or Cr (blue values) solutes. The reference local magnetic moment in bulk Fe is $+2.23\, \mu_\mathrm{B}$.}
\end{figure*}

Classic strain-relief arguments suggest that oversized impurities (Cu, Ni, Mn, Cr) should hold a repulsive interaction in configurations 'M' and '1b', and an attractive one in '1a' -- the opposite for undersized impurities (P, Si) \cite{olsson_ab_2010}. This is indeed true for P, and to a certain extent for Cu, Ni, and Si, but does not apply to Cr and Mn, since the corresponding 'M' and '1b' configurations are stable. Many studies have shown that solute-defect interaction energies cannot be explained based on size-related arguments only and are mainly determined by electronic interactions \cite{janotti_solute_2004, ohnuma_first-principles_2009, olsson_ab_2010, messina_systematic_2016}. In particular, the out-of-trend interactions of Cr and Mn with vacancies and dumbbells are related to electronic and magnetic arguments. Both species hold antiferromagnetic (AFM) moments in a perfect Fe lattice, but the presence of a dumbbell nearby weakens their AFM character and induces a "rearrangement" of the local moments (cf. Fig. \ref{fig:afm_moments}), as it was already mentioned in a previous FeCr study \cite{olsson_ab_2007}. In FeMn, this is consistent with the decrease of magnetic moments occurring next to a foreign interstitial impurity (C, N, O), which was ascribed to an increase of the local charge \cite{schneider_local_2018}. However, the magnetic moments in Fig. \ref{fig:afm_moments} for Cr and Mn follow a similar trend, so the differences in solute-dumbbell binding-energies cannot be explained based on magnetism only.

Migration barriers and trapping configurations are further discussed in Section \ref{sec:discussion_DFT} in relation to RR experiments. In summary, the MD is stable for  P, Mn, and Cr to a lesser extent, and not favorable for Si, Ni, and Cu. The stability of the first three MDs was already known from resistivity recovery (RR) experiments in bcc iron \cite{maury_study_1987, maury_interstitial_1990, abe_interaction_1999} and previous theoretical studies \cite{domain_diffusion_2005, meslin_theoretical_2007, olsson_ab_2007, olsson_ab_2010}, and suggests that these solutes are likely to diffuse via a dumbbell mechanism. However, the MD migration and dissociation rates may play an important role. Furthermore, while Cu and Ni diffusion seems unlikely, the Fe-Si dumbbell has a nearly zero interaction, so it is not possible to tell \textit{a priori} if solute transport takes place. 

\begin{table*}[htb!]
	\caption{DFT solute-dumbbell binding energies (cf. Fig. \ref{fig:configurations}) and migration barriers (cf. Fig. \ref{fig:jump_nomenc}) in eV, compared with previous calculations \cite{vincent_ab_2006, meslin_theoretical_2007, olsson_ab_2009, choudhury_ab-initio_2011} and resistivity-recovery experiments \cite{takaki_resistivity_1983}. Energy barriers for the mixed-dumbbell translation, onsite rotation, and 2nn jump are also reported. The second column refers to Barbe's nomenclature \cite{barbe_phenomenological_2007}.}	
	\renewcommand{\arraystretch}{1.1}	
	\centering
	\begin{adjustbox}{max width=\textwidth}
		\begin{tabular}{C{2cm}C{2cm}|C{1.3cm}C{1.3cm}|C{1.3cm}C{1.3cm}|C{1.3cm}C{1.3cm}|C{1.3cm}C{1.3cm}|C{1.3cm}C{1.3cm}|C{1.3cm}C{1.3cm}}
			\hline
			\hline
			\multicolumn{14}{c}{\addstackgap[4pt]{\textbf{Solute-dumbbell binding energies (positive = attraction)}}} \\
			\hline
			\multicolumn{2}{c|}{{\scriptsize (Fig. \ref{fig:configurations})}}	&	\multicolumn{2}{c|}{ \addstackgap[4pt]{\textbf{P}}}  &	\multicolumn{2}{c|}{\textbf{Mn}}  &	\multicolumn{2}{c|}{\textbf{Cr}}  &	\multicolumn{2}{c|}{\textbf{Si}}  &	\multicolumn{2}{c|}{\textbf{Ni}}  &	\multicolumn{2}{c}{\textbf{Cu}}  \\
			\hline
			\multicolumn{2}{c|}{\scriptsize M} 			&	\multicolumn{2}{c|}{+1.025}												&	\multicolumn{2}{c|}{+0.555}						&	\multicolumn{2}{c|}{+0.045}							&	\multicolumn{2}{c|}{-0.002}							&	\multicolumn{2}{c|}{-0.191}								&	\multicolumn{2}{c}{-0.380}		\\
			\multicolumn{2}{c|}{ \scriptsize 1a, 1b} 	&	\multicolumn{2}{c|}{-0.331, +0.855}									&	\multicolumn{2}{c|}{+0.107, +0.305}		&	\multicolumn{2}{c|}{-0.065, +0.038}				&	\multicolumn{2}{c|}{-0.173, +0.274}				&	\multicolumn{2}{c|}{+0.016, +0.065}					&	\multicolumn{2}{c}{+0.188, +0.065}	\\
			\multicolumn{2}{c|}{\scriptsize 2a, 2b} 	&	\multicolumn{2}{c|}{-0.012, -- $^a$\phantom{0.00}}	&	\multicolumn{2}{c|}{-0.006, +0.083}			&	\multicolumn{2}{c|}{-0.082, -0.081}				&	\multicolumn{2}{c|}{-0.064, +0.045}			&	\multicolumn{2}{c|}{-0.047, +0.027}					&	\multicolumn{2}{c}{+0.099, +0.064}	\\
			\multicolumn{2}{c|}{\scriptsize 3a, 3b, 3c} &	\multicolumn{2}{c|}{-0.05, +0.04, -0.12}					&	\multicolumn{2}{c|}{-0.06, +0.01, +0.01}	&	\multicolumn{2}{c|}{-0.06, -0.03, -0.03}	&	\multicolumn{2}{c|}{-0.01, -0.02, -0.07}	&	\multicolumn{2}{c|}{-0.03, -0.00, -0.04}		&	\multicolumn{2}{c}{+0.02, +0.06, +0.10}	\\
			\multicolumn{2}{c|}{\scriptsize 4a, 4b, 4c} &	\multicolumn{2}{c|}{-0.01, +0.04, -0.02}					&	\multicolumn{2}{c|}{-0.03, +0.02, -0.00}	&	\multicolumn{2}{c|}{-0.04, -0.05, -0.03}	&	\multicolumn{2}{c|}{-0.01, +0.08, +0.02}	&	\multicolumn{2}{c|}{-0.02, +0.03, +0.01}		&	\multicolumn{2}{c}{+0.01, +0.07, +0.05}	\\
			\multicolumn{2}{c|}{\scriptsize 5a, 5b}		&		\multicolumn{2}{c|}{-0.031, +0.212}							&	\multicolumn{2}{c|}{-0.004, +0.013}				&	\multicolumn{2}{c|}{-0.023, -0.033}				&	\multicolumn{2}{c|}{-0.000, +0.038}			&	\multicolumn{2}{c|}{+0.011, -0.024}					&	\multicolumn{2}{c}{+0.052, +0.017}		\\
			\hline
			\multicolumn{14}{c}{\addstackgap[4pt]{\textbf{Jump frequencies}} } \\
			\hline		
			{\scriptsize (Fig.\ref{fig:jump_nomenc})}&	{\scriptsize (Barbe)\textsuperscript{\cite{barbe_phenomenological_2007}} } &	\multicolumn{2}{c|}{\addstackgap[4pt]{\textbf{P}}}  &	\multicolumn{2}{c|}{ \textbf{Mn}}  &	\multicolumn{2}{c|}{ \textbf{Cr}}  &	\multicolumn{2}{c|}{ \textbf{Si}}  &	\multicolumn{2}{c|}{ \textbf{Ni}}  &	\multicolumn{2}{c}{ \textbf{Cu}}  \\
			\hline
			\multicolumn{2}{c|}{{\scriptsize Mixed dumbbell (MD) jumps}}	&	\multicolumn{2}{c|}{}  &	\multicolumn{2}{c|}{}  &	\multicolumn{2}{c|}{}  &	\multicolumn{2}{c|}{}  &	\multicolumn{2}{c|}{}  &	\multicolumn{2}{c}{}  	\\
			$\omega_1$		&		$\omega_1$		&		\multicolumn{2}{c|}{0.217}	&		\multicolumn{2}{c|}{0.316}	&		\multicolumn{2}{c|}{0.241}	&		\multicolumn{2}{c|}{0.568}	&		\multicolumn{2}{c|}{0.464}	&		\multicolumn{2}{c}{0.364}	\\
			$\nu_{\omega_1}$  	&	-- &	\multicolumn{2}{c|}{4.229 {\scriptsize THz}}	&		\multicolumn{2}{c|}{5.930 {\scriptsize THz}}	&		\multicolumn{2}{c|}{4.555 {\scriptsize THz}}	&		\multicolumn{2}{c|}{19.268 {\scriptsize THz}}	&		\multicolumn{2}{c|}{2.828 {\scriptsize THz}}	&		\multicolumn{2}{c}{2.685 {\scriptsize THz}}	\\
			$\tau_1$		&		$\tau_1\omega_1$		&		\multicolumn{2}{c|}{0.493}	&		\multicolumn{2}{c|}{0.648}	&		\multicolumn{2}{c|}{0.407}	&		\multicolumn{2}{c|}{0.575}	&		\multicolumn{2}{c|}{0.634}	&		\multicolumn{2}{c}{0.368}	\\
			$\omega_\mathrm{R1}$		&		$\omega_\mathrm{R1}$		&		\multicolumn{2}{c|}{0.327}	&		\multicolumn{2}{c|}{0.314}	&		\multicolumn{2}{c|}{--}	&		\multicolumn{2}{c|}{--}	&		\multicolumn{2}{c|}{--}	&		\multicolumn{2}{c}{--}	\\
			\multicolumn{2}{c|}{{\scriptsize MD association/dissociation}}	&	 	&		&		&		&		&		&	&		&	&		&		&	 	\\
			$\omega_\mathrm{M\,1b}$, $\omega_\mathrm{1b\,M}$		&	$\omega_2$, $\omega_3$	&	0.230	&	0.060	&	0.448	&	0.198	&	0.365	&	0.358	&	0.069	&	0.345	&	0.083	&	0.339	&	0.000$^b$	&	0.445	\\
			\hline
			\multicolumn{2}{c|}{{\scriptsize Jumps from config. 1b}}	&	\multicolumn{2}{c|}{}  &	\multicolumn{2}{c|}{}  &	\multicolumn{2}{c|}{}  &	\multicolumn{2}{c|}{}  &	\multicolumn{2}{c|}{}  &	\multicolumn{2}{c}{}  	\\
			$\omega_\mathrm{1b2b}$, $\omega_\mathrm{2b1b}$		&	$\omega_4$, $\omega_5$	&	--			&	-- $^c$			&	0.378	&	0.156	&	0.373	&	0.254	&	0.439	&	0.210	&	0.312	&	0.275	&	0.305	&	0.305	\\
			$\omega_\mathrm{1b3b}$, $\omega_\mathrm{3b1b}$		&	$\omega_4$, $\omega_5$	&	0.810	&	0.000$^b$	&	0.521	&	0.227	&	0.386	&	0.318	&	0.475	&	0.184	&	0.396	&	0.327	&	0.304	&	0.302	\\
			$\omega_\mathrm{1b5b}$, $\omega_\mathrm{5b1b}$		&	$\omega_4$, $\omega_5$	&	0.642	&	0.000$^b$	&	0.511	&	0.219	&	0.401	&	0.330	&	0.476	&	0.240	&	0.405	&	0.316	&	0.357	&	0.309	\\
			\multicolumn{2}{c|}{{\scriptsize Jumps from config. 1a}}	&	\multicolumn{2}{c|}{}  &	\multicolumn{2}{c|}{}  &	\multicolumn{2}{c|}{}  &	\multicolumn{2}{c|}{}  &	\multicolumn{2}{c|}{}  &	\multicolumn{2}{c}{}  	\\
			$\omega_\mathrm{1a2a}$, $\omega_\mathrm{2a1a}$		&	$\omega_6$, $\omega_7$	&	0.185	&	0.504	&	0.351	&	0.239	&	0.326	&	0.309	&	0.294	&	0.403	&	0.353	&	0.289	&	0.357	&	0.269	\\
			$\omega_\mathrm{1a2b}$, $\omega_\mathrm{2b1a}$		&	$\omega_6$, $\omega_7$	&	--			&	-- $^c$	&	0.374	&	0.350	&	0.411	&	0.395	&	0.229	&	0.447	&	0.311	&	0.323	&	0.321	&	0.198	\\
			$\omega_\mathrm{1a3b}$, $\omega_\mathrm{3b1a}$		&	$\omega_6$, $\omega_7$	&	0.124	&	0.499	&	0.367	&	0.272	&	0.321	&	0.356	&	0.278	&	0.434	&	0.361	&	0.341	&	0.388	&	0.263	\\
			$\omega_\mathrm{1a3c}$, $\omega_\mathrm{3c1a}$		&	$\omega_6$, $\omega_7$	&	0.283	&	0.490	&	0.368	&	0.270	&	0.352	&	0.383	&	0.312	&	0.411	&	0.366	&	0.315	&	0.359	&	0.269	\\
			\multicolumn{2}{c|}{{\scriptsize Jumps from 2nn}}	&	\multicolumn{2}{c|}{}  &	\multicolumn{2}{c|}{}  &	\multicolumn{2}{c|}{}  &	\multicolumn{2}{c|}{}  &	\multicolumn{2}{c|}{}  &	\multicolumn{2}{c}{}  	\\
			$\omega_\mathrm{2a4c}$, $\omega_\mathrm{4c2a}$		&	--	&	0.438	&	0.432	&	0.377	&	0.382	&	0.309	&	0.362	&	0.388	&	0.470	&	0.303	&	0.358	&	0.352	&	0.302	\\
			$\omega_\mathrm{2b4b}$, $\omega_\mathrm{4b2b}$		&	--	&	--	&	-- $^c$	&	0.334	&	0.274	&	0.314	&	0.349	&	0.289	&	0.325	&	0.289	&	0.288	&	0.256	&	0.265	\\
			$\omega_\mathrm{2b4c}$, $\omega_\mathrm{4c2b}$		&	--	&	--	&	-- $^c$	&	0.376	&	0.293	&	0.314	&	0.366	&	0.362	&	0.335	&	0.336	&	0.316	&	0.302	&	0.286	\\
			\hline
			\multicolumn{14}{c}{\addstackgap[4pt]{ \textbf{Previous calculations}} } \\
			\hline		
			$\omega_1$		&		$\omega_1$		&		\multicolumn{2}{c|}{\addstackgap[4pt]{0.18\textsuperscript{\cite{meslin_theoretical_2007}}}}	&		\multicolumn{2}{c|}{0.34\textsuperscript{\cite{vincent_ab_2006}}}	&		\multicolumn{2}{c|}{0.23\textsuperscript{\cite{olsson_ab_2009}} 0.25\textsuperscript{\cite{choudhury_ab-initio_2011}}}	&		\multicolumn{2}{c|}{0.52\textsuperscript{\cite{vincent_ab_2006}}}	&		\multicolumn{2}{c|}{0.41\textsuperscript{\cite{choudhury_ab-initio_2011}} 0.41\textsuperscript{\cite{vincent_ab_2006}}}	&		\multicolumn{2}{c}{0.32\textsuperscript{\cite{vincent_ab_2006}}}	\\
			$\tau_1$		&		$\tau_1\omega_1$		&		\multicolumn{2}{c|}{0.24\textsuperscript{\cite{meslin_theoretical_2007}}}	&		\multicolumn{2}{c|}{0.66\textsuperscript{\cite{vincent_ab_2006}}}	&		\multicolumn{2}{c|}{0.42\textsuperscript{\cite{olsson_ab_2009}} 0.48\textsuperscript{\cite{choudhury_ab-initio_2011}}}	&		\multicolumn{2}{c|}{0.37\textsuperscript{\cite{vincent_ab_2006}}}	&		\multicolumn{2}{c|}{0.69\textsuperscript{\cite{choudhury_ab-initio_2011}} 0.46\textsuperscript{\cite{vincent_ab_2006}}}	&		\multicolumn{2}{c}{0.26\textsuperscript{\cite{vincent_ab_2006}}}	\\
			$\omega_\mathrm{R1}$		&		$\omega_\mathrm{R1}$		&		\multicolumn{2}{c|}{0.24\textsuperscript{\cite{meslin_theoretical_2007}}}	&		\multicolumn{2}{c|}{0.45\textsuperscript{\cite{vincent_ab_2006}}}	&		\multicolumn{2}{c|}{0.36\textsuperscript{\cite{olsson_ab_2009}}}	&		\multicolumn{2}{c|}{0.48\textsuperscript{\cite{vincent_ab_2006}}}	&		\multicolumn{2}{c|}{0.36\textsuperscript{\cite{vincent_ab_2006}}}	&		\multicolumn{2}{c}{0.32\textsuperscript{\cite{vincent_ab_2006}}}	\\		
			$\omega_\mathrm{1(2nn)}$		&		--		&		\multicolumn{2}{c|}{0.18\textsuperscript{\cite{meslin_theoretical_2007}}}	&		\multicolumn{2}{c|}{0.53\textsuperscript{\cite{vincent_ab_2006}}}	&		\multicolumn{2}{c|}{0.43\textsuperscript{\cite{olsson_ab_2009}}}	&		\multicolumn{2}{c|}{0.67\textsuperscript{\cite{vincent_ab_2006}}}	&		\multicolumn{2}{c|}{0.84\textsuperscript{\cite{vincent_ab_2006}}}	&		\multicolumn{2}{c}{0.59\textsuperscript{\cite{vincent_ab_2006}}}	\\
			$\omega_\mathrm{M\,1b}$, $\omega_\mathrm{1b\,M}$		&	$\omega_2$, $\omega_3$	&	\multicolumn{2}{c|}{1.26, 0.34\textsuperscript{\cite{meslin_theoretical_2007}}}	&\multicolumn{2}{c|}{0.49, 0.22\textsuperscript{\cite{vincent_ab_2006}}}	&\multicolumn{2}{c|}{0.33, 0.30\textsuperscript{\cite{olsson_ab_2009}}}	&\multicolumn{2}{c|}{0.06, 0.35\textsuperscript{\cite{vincent_ab_2006}}}	&\multicolumn{2}{c|}{0.09, 0.33\textsuperscript{\cite{choudhury_ab-initio_2011}}}	&\multicolumn{2}{c}{0.00, 0.50\textsuperscript{\cite{vincent_ab_2006}}}	\\
			$\omega_\mathrm{M\,2b}$, $\omega_\mathrm{2b\,M}$		&	--	&	--	&	--	&\multicolumn{2}{c|}{0.80, 0.04\textsuperscript{\cite{vincent_ab_2006}}}	&\multicolumn{2}{c|}{0.52, 0.36\textsuperscript{\cite{olsson_ab_2009}}}	&\multicolumn{2}{c|}{0.26, 0.14\textsuperscript{\cite{vincent_ab_2006}}}	&\multicolumn{2}{c|}{0.23, 0.45\textsuperscript{\cite{vincent_ab_2006}}}	&\multicolumn{2}{c}{0.05, 0.33\textsuperscript{\cite{vincent_ab_2006}}}	\\		
			$\omega_\mathrm{1b2b}$, $\omega_\mathrm{2b1b}$		&	$\omega_4$, $\omega_5$	&	--	&	--	&	--	&	--	&	\multicolumn{2}{c|}{0.35, 0.22\textsuperscript{\cite{olsson_ab_2009}}}	&	--	&	--	&	\multicolumn{2}{c|}{0.31, 0.27\textsuperscript{\cite{choudhury_ab-initio_2011}}}	&	--	&	--	\\
			$\omega_\mathrm{1a2b}$, $\omega_\mathrm{2b1a}$		&	$\omega_6$, $\omega_7$	&	--	&	--	&	--	&	--	&	\multicolumn{2}{c|}{0.36, 0.37\textsuperscript{\cite{choudhury_ab-initio_2011}}}	&	--	&	--	&	\multicolumn{2}{c|}{0.36, 0.34\textsuperscript{\cite{choudhury_ab-initio_2011}}}	&	--	&	--	\\
			\hline
			\hline
			\multicolumn{14}{l}{\scriptsize $^a$ Configuration 2b in Fe(P) is unstable, as it relaxes into a mixed dumbbell.} \\
			\multicolumn{14}{l}{\scriptsize $^b$ Athermal jump with no energy barrier.} \\
			\multicolumn{14}{l}{\scriptsize $^c$ Not computed because of the instability of the initial or final configuration.} \\
		\end{tabular}
	\end{adjustbox}
	\label{tab:DFT_results}
\end{table*}

\subsection{Dumbbell attempt frequencies}

\begin{figure}[htb!]
	\centering
	\resizebox{0.6\columnwidth}{!}{
		\includegraphics{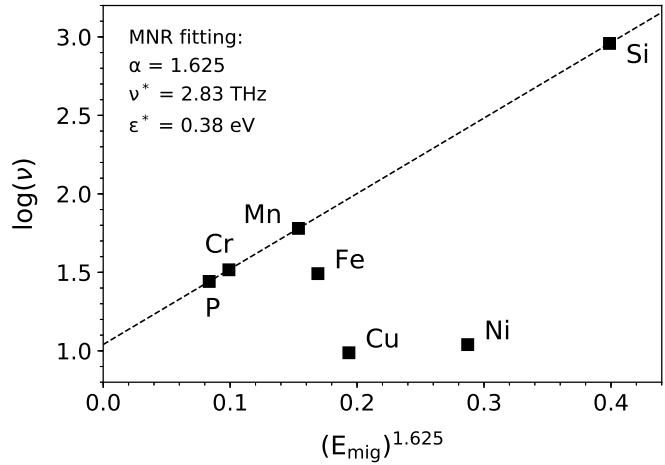}}
	\caption{\label{fig:MNR_fitting} Fitting of the pure- and mixed-dumbbell migration energies and attempt frequencies according to the Meyer-Neldel rule (Eq. \ref{eq:mnr_rule}), achieved by excluding the Fe, Cu, and Ni data points from the fitting.}
\end{figure}

The RT attempt frequencies of the mixed dumbbells are listed in Table \ref{tab:DFT_results}. The values vary in a small range around the attempt frequency in pure Fe, with the exception of Si. According to the Meyer-Neldel compensation rule (MNR) \cite{marinica_diffusion_2005}, the prefactors $\nu$ should be correlated to the corresponding migration barriers $E^\mathrm{mig}$ as:
\begin{equation}
\ln \left(\frac{\nu}{\nu^*} \right) = \left( \frac{E^\mathrm{mig}}{\varepsilon^*} \right)^\alpha \,
\label{eq:mnr_rule}
\end{equation}
where the parameters $\nu^*$, $\varepsilon^*$, and $\alpha$ depend on the diffusion process, and provide information about the type of vibrational modes involved in the transition. The prefactors should thus tend to be higher for high-barrier jumps, and viceversa. This rule was successfully applied to vacancy-solute exchange in Fe \cite{messina_systematic_2016} in spite of the missing multi-phonon excitations in TST harmonic calculations. In Fig. \ref{fig:MNR_fitting}, Eq. \ref{eq:mnr_rule} is applied to the MD jumps, and a perfect correlation ($R^2 = 1$) is found for P, Cr, Mn, and Si with an exponent $\alpha=1.625$ and fitting parameters $\nu^* = 2.830$ THz and $\varepsilon^*=0.38$ eV. Fe is also not far from the MNR fit. Such a high $\alpha$ exponent might suggest that the migration is guided by optical modes, as was the case for 5d elements exchanging with a vacancy in Fe. On the other hand, Ni and Cu do not fulfill this condition, so different vibrational phenomena might be at play. Further investigations beyond the harmonic approximation \cite{swinburne_unsupervised_2018} are needed to clarify this anomaly.


\begin{figure}[htb!]
	\centering
	\resizebox{0.6\columnwidth}{!}{
		\includegraphics{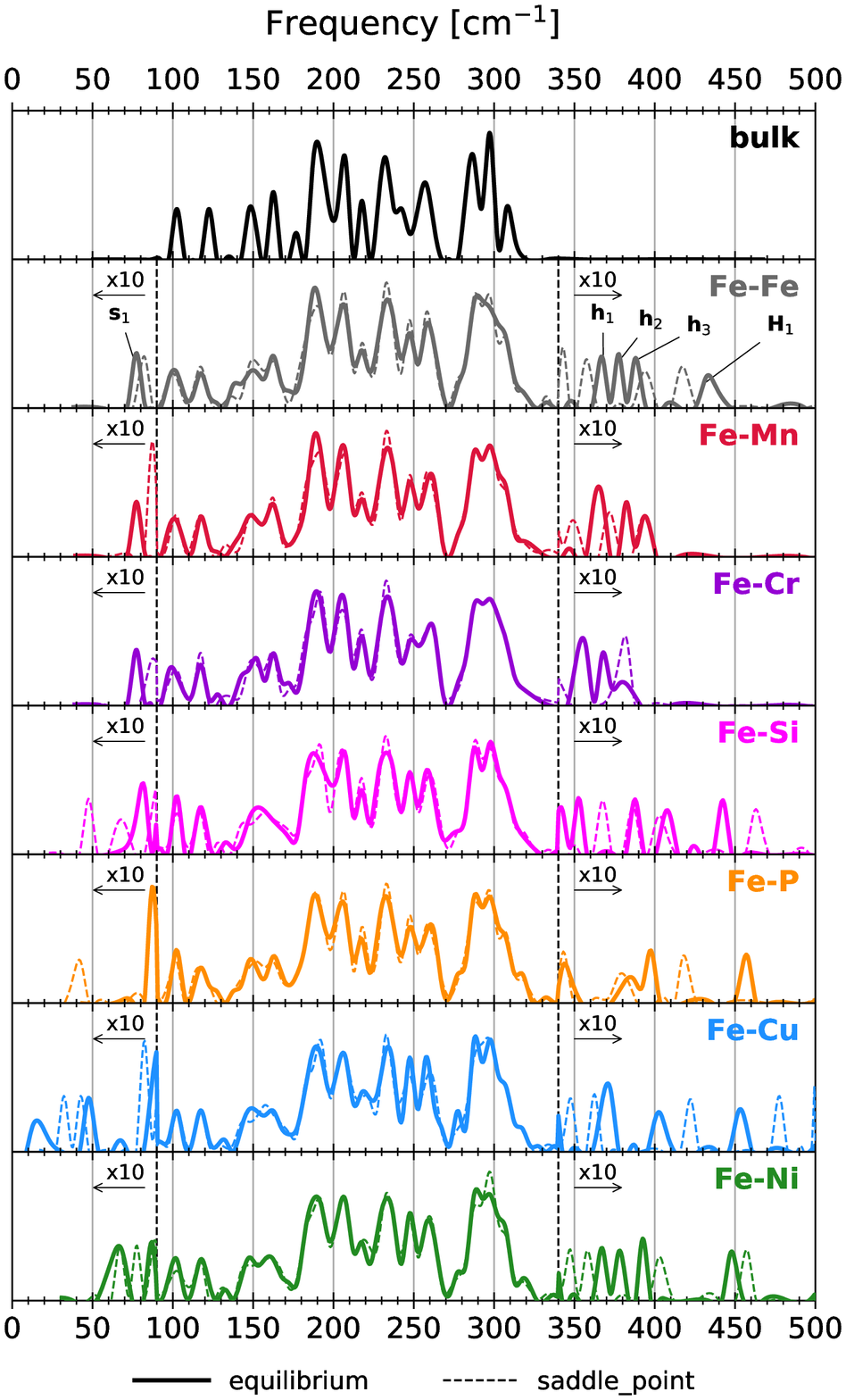}}
	\caption{\label{fig:phonon_spectra}DFT-computed phonon spectra for the Fe-Fe and Fe-$X$ dumbbells in the equilibrium configuration (continuous lines) and at the saddle point (dashed lines) of the rotation-translation jump. Areas marked with 'x10' indicate that the spectrum magnitude is amplified by 10 times. }
\end{figure}

Figure \ref{fig:phonon_spectra} shows the phonon spectra for each dumbbell in the equilibrium configuration and at the saddle point. The equilibrium spectrum of the $\langle110 \rangle$ pure dumbbell is in good agreement with Lucas \textit{et al.} \cite{lucas_vibrational_2009}. With respect to the bulk spectrum, it contains a soft mode (s1) and some hard modes (h1, h2, h3, and H). According to Lucas, the hard modes are related to the stretching of the dumbbell bond (H) or the surrounding ones (h1, h2, h3) (their higher frequency is due to bond compression). The soft mode is instead associated to a translation of the dumbbell along the $\langle110 \rangle$ direction, possibly favoring the RT mechanism.

Except Fe-Cr and Fe-Mn, the mixed dumbbells have similar high phonon modes, reaching up to 500 THz (Fe-Ni). Likewise in pure Fe-Fe, these high modes are due to the stretching of the bonds. The Fe-Cu and Fe-Ni dumbbells have very low frequency modes that compensate the strongly negative binding energies, whereas for all the other solutes, with positive binding energies, these instabilities are missing. The pronounced soft modes in Fe-Cu and Fe-Ni induce a flat energetic landscape and can stabilize these mixed dumbbells at higher temperatures. At the saddle point, the Fe-Cu and Fe-Ni dumbbells have the same low-frequency modes and flat energetic landscape. The migration pathway for Cu and Ni is thus probably related to low-frequency modes, which might explain the mismatch with the MNR rule for these elements.


\section{Solute transport}
\label{sec:solute_transport}

The DFT data presented in the previous section is used as input in the numerical part of KineCluE to obtain the vacancy- and dumbbell transport coefficients for each Fe-\textit{X} binary alloy, applying the kinetic cluster-expansion approach of Eq. \ref{eq:cluster_expansion}. The complete set of results is available in the associated database \cite{database}. Solute-defect flux coupling and transport is then analyzed based on the following quantities:
\begin{itemize}
	\item[-] The flux-coupling ratios involving the off-diagonal coefficients: $L_\mathrm{V B}^\mathrm{(V B)} / L_\mathrm{BB}^\mathrm{(V B)}$ and $L_\mathrm{\delta B}^\mathrm{(\delta B)} / L_\mathrm{\delta \delta}^\mathrm{(\delta B)}$ ($\delta=\mathrm{V, I}$), which determine the mutual directions of defect and solute fluxes.
	\item[-] The solute tracer diffusion coefficients $D_{\mathrm{B}^*}^\delta$ obtained from the solute diagonal coefficients $L_\mathrm{BB}^\mathrm{(\delta B)}$; the comparison between $D_{\mathrm{B}^*}^\mathrm{V}$ and $D_{\mathrm{B}^*}^\mathrm{I}$ reveals the preferential diffusion mechanism for each solute.
	\item[-] The ratios of partial-diffusion coefficients $D_\mathrm{pd}^\delta$, describing the diffusion speed (and direction) of solute atoms relative to matrix atoms, thus the RIS tendencies induced by each mechanism.
\end{itemize}
The mathematical framework used to derive these quantities from the KineCluE output is presented in \ref{sec:appendix_computation_total_lij} and summarized in Table \ref{tab:formula_summary}. Even though the total concentrations of defects $C_\mathrm{V}$, $C_\mathrm{I}$ and solutes $C_\mathrm{B}$ are required as additional parameters, $G^\delta$ and $D_{\mathrm{B}^*}^\delta$ are independent of $C_\mathrm{B}$, and so is $D_\mathrm{pd}^\delta$ for sufficiently low concentrations. The total defect concentrations appear in $D_{\mathrm{B}^*}^\delta$ but not in $G^\delta$ nor $D_\mathrm{pd}^\delta$.

\begin{table*}[!htb]\scriptsize
	\caption{Summary of solute (B) transport by vacancies (V) and dumbbells (I), solute-defect cluster properties, and radiation-induced segregation (RIS) results. The diffusion coefficients, cluster mobility and lifetimes, and the solute-vacancy correlation factors are shown in terms of activation energy and prefactor, after Arrhenius interpolation between 300 and 1000 K. The solute diffusion coefficients are proportional to the corresponding defect concentration, which is here set to 1. In this dilute-limit model, the RIS switchover temperatures are nearly independent of solute concentration $C_\mathrm{B}$, except that of Cr that can vary by $\pm 30$ K with respect to the shown value obtained at $C_\mathrm{B}=0.1$ at.\%.}
	\renewcommand{\arraystretch}{1.6}	
	\centering
	\begin{adjustbox}{width=\textwidth}
		\begin{tabular}{L{2.8cm}L{1.8cm}C{2cm}C{2cm}C{2cm}C{2cm}C{2cm}C{2cm}}
			\hline
			\hline
			\scriptsize
			&  &  \textbf{Cr}  &  \textbf{Cu}  &  \textbf{Mn}  &  \textbf{Ni}  &  \textbf{P}  &  \textbf{Si}  \\
			\hline
			\multicolumn{8}{l}{\textbf{Vacancy-assisted diffusion}}  \\
			\hline
			\multirow{2}{2.8cm}{B diffusion coeff.} &  \scriptsize {$E_\mathrm{act}^\mathrm{V} $ [eV]} &  0.632  & 0.462  &  0.508  &  0.527 &  0.321 &  0.453   \\
			&  \scriptsize{$D_{\mathrm{B}_0}^\mathrm{V} $ [m$^2$/s]}  &  $1.31 \cdot 10^{-6}$  &  $1.14 \cdot 10^{-6}$ &   $1.34 \cdot 10^{-6}$   &  $5.15 \cdot 10^{-7}$  &   $7.66 \cdot 10^{-7}$  &  $8.70 \cdot 10^{-7}$  \\
			\hline
			\multirow{3}{2.8cm}{B correlation factor} 	&  (300-1000 K) & 0.012 -- 0.37 & 0.0012 -- 0.29 & 0.00034 -- 0.19 & 0.89 -- 0.73 & 0.00010 -- 0.17 & 0.00055 -- 0.23 \\
			&  \scriptsize {$E_{\mathrm{F}_\mathrm{B}}^\mathrm{V} $ [eV]} &  0.124  & 0.205  &  0.235  &  -0.008 &  0.276 &  0.223   \\
			&  \scriptsize{$F_{\mathrm{B}_0}^\mathrm{V}$ }  &  1.68  &  3.30 &   3.00   &  0.665  &   4.35  &  3.14  \\
			\hline
			\multirow{2}{2.8cm}{VB pair mobility } &  \scriptsize {$E_\mathrm{mig}^\mathrm{(VB)} $ [eV]} &  0.650  & 0.693  &  0.623  &  0.700 &  0.668 &  0.720   \\
			&  \scriptsize{$M_0^\mathrm{(VB)}$  [m$^2$/s]}  &  $2.12 \cdot 10^{-8}$  &  $6.58 \cdot 10^{-8}$ &   $4.32 \cdot 10^{-8}$   &  $2.89 \cdot 10^{-8}$  &   $6.77 \cdot 10^{-8}$  &  $7.64 \cdot 10^{-8}$  \\
			\hline
			\multirow{2}{2.8cm}{VB pair lifetime } &  \scriptsize {$E_\mathrm{diss}^\mathrm{(VB)} $ [eV]}&  0.706 & 0.908 &  0.805  &  0.853   &  1.042  &  0.965   \\
			&  \scriptsize{$\tau_0^\mathrm{(VB)}$ [s]}   &  $3.24 \cdot 10^{-14}$   &  $1.01 \cdot 10^{-14}$   &   $1.65 \cdot 10^{-14}$   &  $1.08 \cdot 10^{-14}$ &   $6.83 \cdot 10^{-15}$  &  $6.45 \cdot 10^{-15}$   \\
			\hline
			\multicolumn{8}{l}{\textbf{Dumbbell-assisted diffusion}}  \\
			\hline
			\multirow{2}{2.8cm}{B diffusion coeff.} &  \scriptsize {$E_\mathrm{act}^\mathrm{I} $  [eV]} &  0.219  & 0.744  &  -0.231  &  0.654 &  -0.803 &  0.569   \\
			&  \scriptsize{$D_{\mathrm{B}_0}^\mathrm{I} $  [m$^2$/s]}  &  $5.21 \cdot 10^{-7}$  &  $1.29 \cdot 10^{-6}$ &   $4.12 \cdot 10^{-7}$   &  $1.34 \cdot 10^{-6}$  &   $2.39 \cdot 10^{-7}$  &  $3.90 \cdot 10^{-6}$  \\
			\hline
			B correlation factor 	&  (300-1000 K) & 0.11 -- 0.20 & 0.99 -- 1.00 & 0.11 -- 0.14 & 0.99 -- 1.00 & 0.19 -- 0.23 & 0.96 -- 1.00 \\
			\hline
			\multirow{2}{2.8cm}{IB pair mobility } &  \scriptsize {$E_\mathrm{mig}^\mathrm{(IB)} $ [eV]} &  0.217  & 0.881  &  0.319  &  0.674 &  0.213 &  0.824  \\
			&  \scriptsize{$M_0^\mathrm{(IB)}$  [m$^2$/s]}  &  $1.85 \cdot 10^{-9}$  &  $9.12 \cdot 10^{-9}$ &   $2.97 \cdot 10^{-8}$   &  $4.61 \cdot 10^{-9}$  &   $3.07 \cdot 10^{-8}$  &  $2.20 \cdot 10^{-7}$  \\
			\hline
			\multirow{2}{2.8cm}{IB pair lifetime } &  \scriptsize {$E_\mathrm{diss}^\mathrm{(IB)} $ [eV]}&  0.348 & 0.448 &  0.884  &  0.355   &  1.290  &  0.579  \\
			&  \scriptsize{$\tau_0^\mathrm{(IB)}$ [s]}   &  $6.72 \cdot 10^{-14}$   &  $3.64 \cdot 10^{-14}$   &   $3.45 \cdot 10^{-15}$   &  $7.30 \cdot 10^{-14}$ &   $7.20 \cdot 10^{-15}$  &  $1.08 \cdot 10^{-14}$   \\
			\hline
			\multicolumn{2}{l}{\textbf{Dominant diffusion mechanism}}								& dumbbell	&  vacancy	&  dumbbell & vacancy/ both  &  dumbbell  &  both  \\
			\hline
			\multicolumn{2}{l}{\textbf{Vacancy drag max temperature}} 					&	243 K &  1085 K & 	 997 K & 	 1074 K & 	 2060 K & 	 1308 K   \\
			\hline
			\multicolumn{2}{l}{\textbf{RIS switchover temperature}}								& 543 $\pm 30$ K	&  1086 K	&  enrichment &  1089 K  &  enrichment  & enrichment  \\
			\hline
			\multicolumn{2}{l}{RIS tendency by dumbbells}  & enrichment & (negligible) &  enrichment & depletion & enrichment & depl./enrich. \\
			\hline
			\multicolumn{2}{l}{RIS tendency by vacancies}  & depletion & enrichment  & enrichment & enrichment & enrichment & enrichment \\
			& & & for $T<1086 $ K  & for $T<1084 $ K  & for $T<2218 $ K  &   & for $T<1606 $ K  \\
			\hline
			\hline
		\end{tabular}
	\end{adjustbox}
	\label{tab:results_recap}
\end{table*}

\subsection{Properties of solute-defect pairs}
\label{sec:transport_clusters}

In addition to transport coefficients, KineCluE yields cluster properties that can be of direct use in coarse-grained methods such as object Kinetic Monte Carlo (OKMC) or cluster dynamics: the average mobility, lifetime, and mean free path (MFP) before dissociation. To this purpose, the kinetic radius is set to a smaller value that corresponds to the cluster cut-off radius, i.e., the mutual distance of atoms beyond which the cluster is considered as dissociated. The chosen cut-off radius is $\sqrt{3}\;a_0$ (5nn distance). Following the theoretical framework explained in a previous work \cite{schuler_transport_2016}, each cluster transport coefficient is expressed as a sum of two contributions related respectively to mobility and association-dissociation jumps. Mobility is obtained by zeroing all dissociative jump frequencies. The dissociation rate is equal to the sum of the dissociation frequencies, multiplied by the total probability of being in a configuration from which the cluster can dissociate. The lifetime is then the inverse of the dissociation rate. Both quantities can be fitted with an Arrhenius curve as:
\begin{equation}
M^\mathrm{(\delta B)} = M_0^\mathrm{(\delta B)} \exp \left( - \frac{E_\mathrm{mig}^\mathrm{(\delta B)}}{k_\mathrm{B}T}  \right) \;\; ,
\end{equation}
and
\begin{equation}
\tau^\mathrm{(\delta B)} = \tau_0^\mathrm{(\delta B)} \exp \left( \frac{E_\mathrm{diss}^\mathrm{(\delta B)}}{k_\mathrm{B}T}  \right) \;\; ,
\end{equation}
where $E_\mathrm{mig}^\mathrm{(\delta B)}$ and $E_\mathrm{diss}^\mathrm{(\delta B)}$ are respectively the pair migration and dissociation energy, and $M_0^\mathrm{(\delta B)}$, $\tau_0^\mathrm{(\delta B)}$ the associated prefactors. The MFP for 3-dimensional migration is then given by $\Delta R = \sqrt{6M \tau}$.

\begin{figure}[!htb]
	\centering
	\resizebox{0.6\columnwidth}{!}{
		\includegraphics{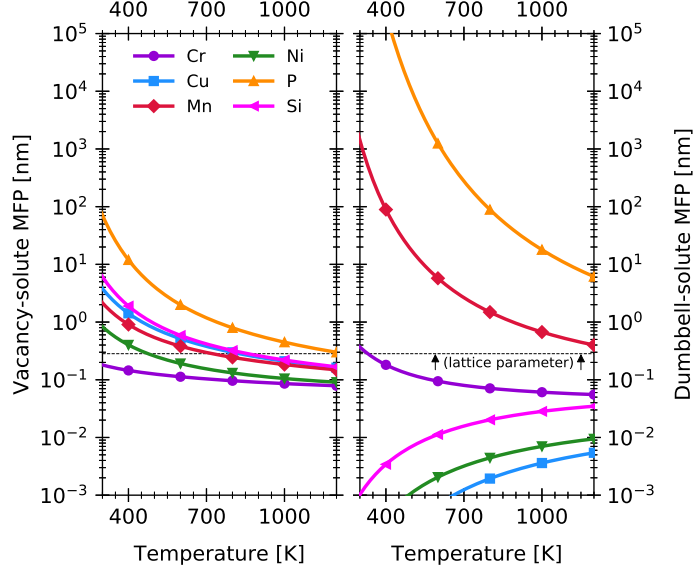}}
	\caption{\label{fig:mfp} Mean free paths (MFP) of vacancy- (left) and dumbbell-(right) solute pairs, computed in KineCluE with a cluster cut-off radius of $\sqrt{3}\;a_0$ (5nn distance).}
\end{figure}

The results of the Arrhenius fitting in the range 300-1000 K are reported in Table \ref{tab:results_recap}, and the MFPs in Fig. \ref{fig:mfp} as functions of temperature. The results compare qualitatively well with previous AKMC calculations of vacancy-solute properties, performed with the same DFT migration barriers but different attempt frequencies and under slightly different assumptions \cite{messina_stability_2015}. In this sense, KineCluE represents a computationally efficient alternative method to AKMC to parameterize coarse-grained simulations, which can be especially useful at low temperatures and in the presence of trapping configurations.

The mobility term represents the kinetic properties of the cluster independently of its stability. As such, it stems from the combination of the solute-defect jump rate ($\omega_2$ for vacancies, $\omega_1$ for dumbbells) with the other jump frequencies comprised  in the kinetic radius. For vacancy-solute pairs, many jump frequencies contribute to the cluster mobility because the $\omega_2$ jump alone is not sufficient to produce a net cluster displacement. For dumbbells, this is not true because the RT jump alone can yield an actual cluster jump. For this reason, the dumbbell-solute mobilities for Cr, Mn, and P are equal to the migration barriers of the $\omega_1$ jumps.

The dissociation energy is the sum of the cluster average binding energy and the PD migration energy in the periphery of the cluster (very close thus to $\omega_0$). Table \ref{tab:results_recap} shows that $E_\mathrm{diss} > E_\mathrm{mig}$ for all stable pairs, i.e., all solute-vacancy pairs and the mixed Fe-Cr, Fe-Mn, and Fe-P dumbbells. On the contrary, $E_\mathrm{mig} > E_\mathrm{diss}$ for the remaining mixed dumbbells, which are thus more likely to dissociate than migrate. This reflects directly on their MFPs that are much shorter than the 1nn distance. The Fe-Cu, Fe-Ni, and Fe-Si dumbbells can be therefore regarded as unstable and sessile. Instead, the MFPs of the Fe-P and Fe-Mn dumbbells are remarkably longer than the corresponding vacancy pairs thanks to the combination of high dissociation and low migration barriers, and can reach distances of the order of $\mu$m and higher. The vacancy-solute MFPs are as long as 1 nm or shorter between 500 and 700 K, except the vacancy-P pair whose MFP is roughly one order of magnitude higher. Finally, the MFPs of pairs containing Cr atoms are rather limited for both mechanisms.

\subsection{Solute diffusion coefficients}

\begin{figure}[!htb]
	\centering
	\resizebox{0.6\columnwidth}{!}{
		\includegraphics{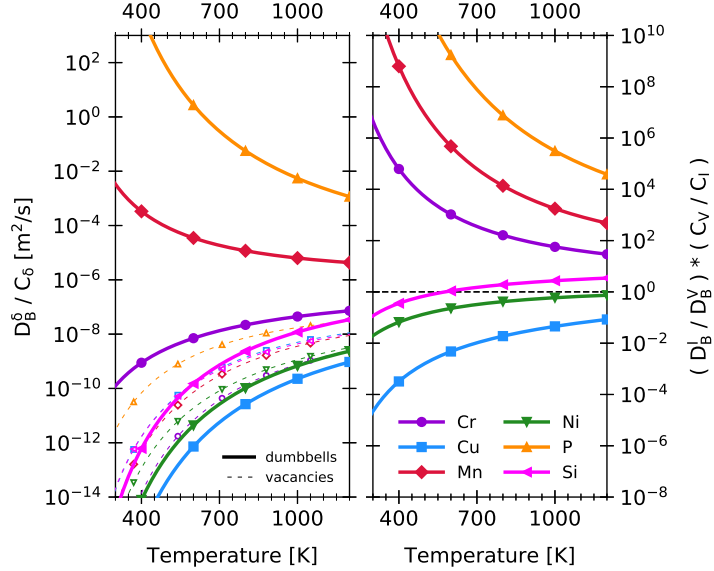}}
	\caption{\label{fig:solute_diff} (left) Solute tracer diffusion coefficients for the dumbbell mechanism (continuous lines) and the vacancy mechanism (dashed lines), divided by the corresponding defect concentration. (right) Ratio between the diffusion coefficients of the two mechanisms, showing that P, Mn, Cr are transported preferentially by dumbbells, and Cu by vacancies.}
\end{figure}

The solute tracer diffusion coefficients, normalized to the corresponding defect concentration, are shown on the left-hand side of Fig. \ref{fig:solute_diff}. 
The ratio of the coefficients stemming from each mechanism is reported on the right-hand side of the same figure. In addition, Table \ref{tab:results_recap} compiles the solute diffusion activation energies and prefactors, fitted in the range 300-1000 K according to a classic Arrhenius fit: 
\begin{equation}
D_{\mathrm{B}}^\delta = D_{\mathrm{B}_0}^\delta \exp \left( - \frac{E_\mathrm{act}^\delta}{k_\mathrm{B}T}\right)  \; .
\label{eq:diffusion_arrhenius}
\end{equation}
The same table shows the solute correlation factor $F_\mathrm{B}^\delta$, which accounts for the solute slowdown (with respect to a random walk) due to exchanges with the defects yielding no net displacement. $F_\mathrm{B}^\mathrm{V}$ follows an Arrhenius behavior analogous to Eq. \ref{eq:diffusion_arrhenius}, and is therefore reported with a fitted activation energy and prefactor, while $F_\mathrm{B}^\mathrm{I}$ is found to be rather independent of temperature. 

The diffusion activation energy is roughly given by  $E_\mathrm{act} = E_\mathrm{mig} - E_\mathrm{bind} + E_\mathrm{F_\mathrm{B}}$, where $E_\mathrm{mig}$ is the solute migration energy ($\omega_2$ for vacancies \cite{messina_exact_2014} and $\omega_1$ for dumbbells). An attractive (positive) solute-defect binding ($E_\mathrm{bind}$) decreases the energy barrier, while correlations ($E_\mathrm{F_\mathrm{B}}$) increase it. In thermal-equilibrium conditions, the defect formation energy should be added to the activation energies shown in Table \ref{tab:results_recap}. This does not apply in irradiation conditions where the defect population is fixed.

Figure \ref{fig:solute_diff} shows that P is the fastest diffuser for both mechanisms. For vacancy diffusion, Cu, Mn, and Si have similar diffusivities, while Ni and Cr are the slowest species. On the other hand, dumbbell coefficients are very high for Mn, very low for Cu and Ni, and in an intermediate range for Cr and Si. It is noteworthy that the fitted energy for $D_\mathrm{P}^\mathrm{I}$  and $D_\mathrm{Mn}^\mathrm{I}$ is negative, which leads to the counter intuitive conclusion that, if the dumbbell formation energy is neglected, P and Mn diffuse faster at low temperature. This stems from the comparison between the MD migration and binding energies: since the dumbbell correlation factor is quite weak (about 0.1-0.2 for P, Mn, Cr, and around unity for Cu, Ni, Si), $E_\mathrm{act}$ is roughly equal to $E_\mathrm{mig} - E_\mathrm{bind}$, and for Fe-Mn and Fe-P dumbbells $E_\mathrm{bind}>E_\mathrm{mig}$. In other words, with decreasing temperatures, the mobility of single P and Mn atoms decreases, but the total diffusivity rises because of the increasing population of mixed dumbbells. 

Owing to weak correlations and lower migration barriers than vacancies, dumbbell diffusion is faster than vacancies for those solutes whose MD is stable. On the right-hand side of Fig. \ref{fig:solute_diff}, it can be seen that P, Mn, and Cr diffuse preferentially by dumbbells, while Cu diffuses exclusively by vacancies. On the other hand, the two mechanisms are surprisingly in competition for Ni and Si atoms, although the balance depends as well on the ratio of PD concentrations $C_\mathrm{V}/C_\mathrm{I}$, which can span a large range of values depending on irradiation and microstructure conditions. Despite the lack of MD stability, and the high $\omega_1$ migration barrier, Si can diffuse by both mechanisms. The compensation of the high $\omega_1$ attempt frequency has thus an important impact on diffusion. For Ni, the MD has a strong repulsive interaction, and yet the vacancy diffusion coefficient is just 3 to 10 times larger than the dumbbell one between 450 and 700 K. Dumbbell diffusion can therefore be as important as vacancy even if the MDs are unstable, not mobile and have short MFPs. Even though the dissociation rate of a single MD is high, the probability of associating with another incoming dumbbell and performing some jumps is still comparable to the energy barriers involved in vacancy diffusion. Ni and Si interstitial diffusion should thus not be neglected in the interpretation of experiments and in microstructure evolution models.

\subsection{Flux-coupling ratios}

\begin{figure}[htb!]
	\centering
	\resizebox{0.6\columnwidth}{!}{
		\includegraphics{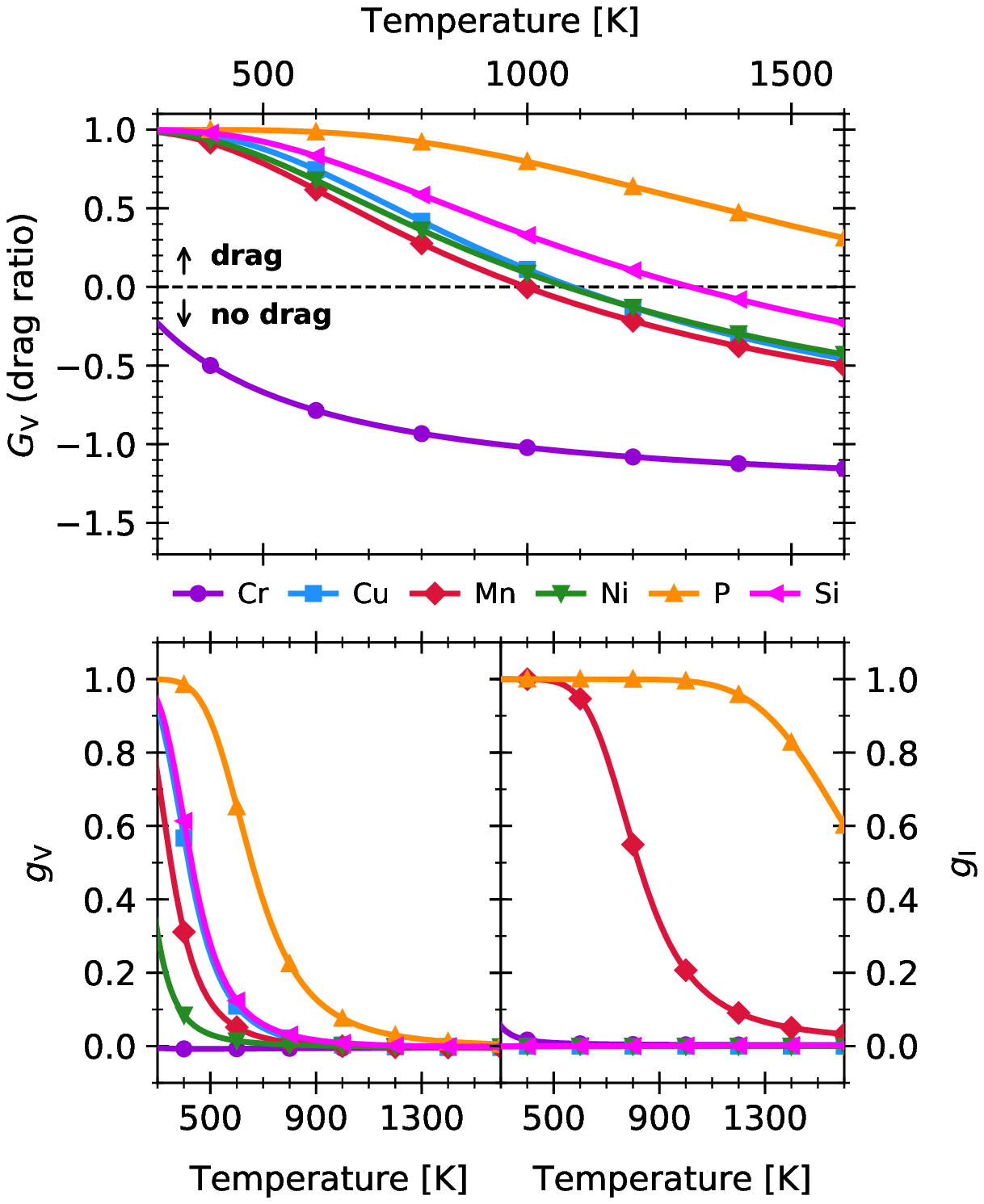}}
	\caption{Above, vacancy drag ratio $G_\mathrm{V}=L_\mathrm{VB}^\mathrm{(VB)}/L_{\mathrm{BB}}^\mathrm{(VB)}$ as a function of temperature. Vacancy drag occurs for positive values of $G_\mathrm{V}$. Below, ratio of the off-diagonal coefficients $g_\mathrm{V} = L_\mathrm{VB}^\mathrm{(VB)}/L_\mathrm{VV}^\mathrm{(VB)}$ and $g_\mathrm{I}=L_\mathrm{IB}^\mathrm{(IB)}/L_\mathrm{II}^\mathrm{(IB)}$ with respect to the vacancy (left) and dumbbell (right) transport coefficient, respectively. The latter can be interpreted as the probability for a defect jump to produce a solute displacement.} 
	\label{fig:drag_factors}
\end{figure}

The flux-coupling ratios are shown in Fig. \ref{fig:drag_factors}. Above, the "conventional" vacancy drag ratio $G_\mathrm{V}=L_\mathrm{VB}^\mathrm{(VB)}/L_{\mathrm{BB}}^\mathrm{(VB)}$ indicates if solute and vacancies diffuse in the same ($G_\mathrm{V}>0$) or opposite ($G_\mathrm{V}<0$) direction. Below, the off-diagonal coefficient ratio with respect to the PD transport coefficient  $g_\delta = L_\mathrm{\delta B}^\mathrm{(\delta B)}/L_\mathrm{\delta\delta}^\mathrm{(\delta B)}$ can be interpreted as the probability for a given diffusing PD to be coupled to a solute atom, or in other words, as the fraction of defect jumps causing a correlated solute jump. Note that the $L_\mathrm{VB}^\mathrm{(VB)}$ coefficient can switch sign, but when negative, the $g_\mathrm{V}$ ratio remains close to zero. On the other hand, the  $L_\mathrm{IB}^\mathrm{(IB)}$ coefficient is always positive because a flux of dumbbells cannot induce a solute flux in the opposite direction. 

As discussed in a previous work \cite{messina_exact_2014}, vacancy drag occurs systematically below a solute-dependent temperature threshold directly related to the extent of solute-vacancy binding. The coupling progressively fades with increasing temperature. At the temperatures of interest for ferritic steels, including RPV steels, all solutes here investigated except Cr are expected to diffuse by vacancy drag. The slightly different set of DFT data and SCMF model here used with respect to the previous study did not lead to any substantial difference. Concerning dumbbell diffusion, the $g_\mathrm{I}$ ratio shows that the defect is strongly correlated to P and Mn atoms, which means that a (mixed) dumbbell jump is likely to produce a solute displacement. P and Mn transport by dumbbells is therefore expected to play an important role.

\subsection{Partial diffusion coefficient ratios}

\begin{figure}[htb!]
	\centering
	\resizebox{0.6\columnwidth}{!}{
		\includegraphics{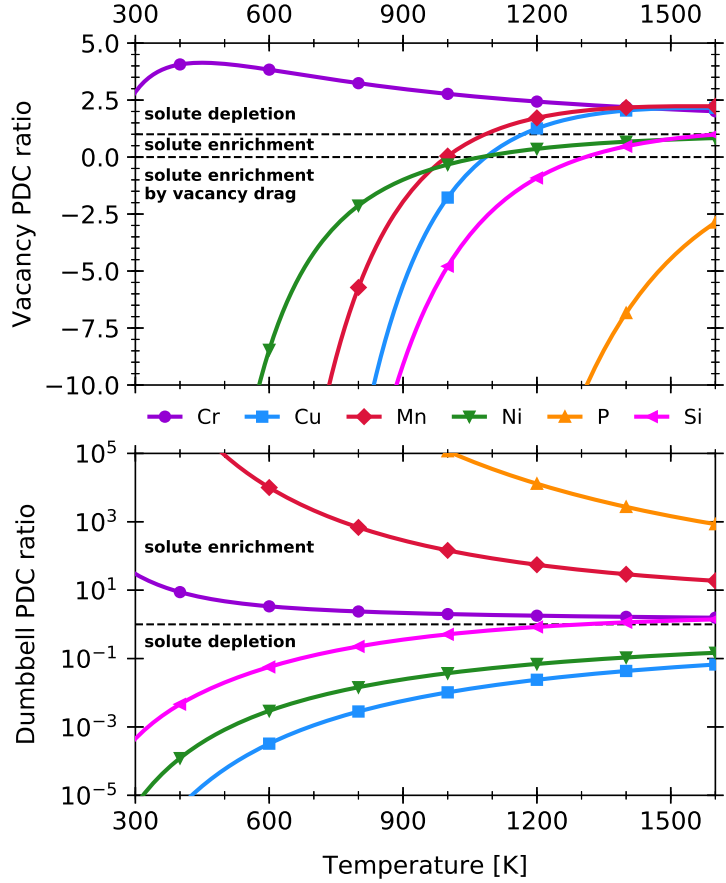}}
	\caption{\label{fig:pdc_ratios} Ratios of partial diffusion coefficients $D^\delta_\mathrm{pd}$ ($\delta = \mathrm{V,I}$) (cf. Table \ref{tab:formula_summary}) for vacancy (top) and dumbbell (bottom) diffusion. Solute enrichment by vacancies occurs via vacancy drag ($D^\mathrm{V}_\mathrm{pd}<0$), or by the inverse Kirkendall mechanism when vacancies exchange preferentially with host atoms ($0 < D^\mathrm{V}_\mathrm{pd} < 1$). Enrichment by dumbbells occurs when $D^\mathrm{I}_\mathrm{pd} > 1$.}
\end{figure}

Figure \ref{fig:pdc_ratios} shows the partial diffusion coefficient (PDC) ratios $D_\mathrm{pd}^\delta$: 
\begin{equation}
	D_\mathrm{pd}^\delta = \frac{\left( 1-C_\mathrm{B} \right)}{C_\mathrm{B}} \cdot \frac{L_\mathrm{\delta B}}{L_\mathrm{\delta A}}
	\label{eq:pdc_ratio}
\end{equation}
(cf. Table \ref{tab:formula_summary}) for vacancy ($\delta=\mathrm{V}$) and dumbbell ($\delta=\mathrm{I}$) diffusion. With respect to the solute diffusion coefficients, the PDC ratios evaluate solute diffusion relative to host atoms, which determines the solute segregation or depletion tendency at sinks. As opposed to the solute-to-solvent diffusion coefficient ratio ($D_\mathrm{B}/D_\mathrm{A}$) commonly used to discuss RIS, the PDC ratio includes flux coupling and is therefore more suitable in systems with correlated fluxes (e.g., in the presence of vacancy drag). 

The sign of the difference $(D_\mathrm{pd}^\mathrm{vac} - D_\mathrm{pd}^\mathrm{sia})$ determines the global RIS tendency (cf. Eq. \ref{eq:alfa_ris}). Therefore, the diffusion mechanism predominantly driving the RIS behavior of each solute can be discussed based on the magnitudes of the two ratios. The contribution of each mechanism can be singled out by setting the PDC ratio of the other mechanism to 1. Consequently, enrichment by vacancies takes place when $D_\mathrm{pd}^\mathrm{vac} < 1$, and by dumbbells when $D_\mathrm{pd}^\mathrm{sia} > 1$. $D_\mathrm{pd}^\mathrm{sia}$ is always positive due to the fact that the dumbbell-solute coupling cannot be negative ($L_\mathrm{IB}>0$), whereas $D_\mathrm{pd}^\mathrm{vac}$ is negative in case of vacancy drag. In the latter case, the difference $(D_\mathrm{pd}^\mathrm{vac} - D_\mathrm{pd}^\mathrm{sia})$ is always negative, which means that an enrichment tendency due to vacancy drag cannot be overturned by a dumbbell-induced depletion.

The vacancy PDC ratios are unchanged with respect to the previous study \cite{messina_exact_2014}. Enrichment by vacancies is expected for all solutes but Cr due to vacancy drag up to high temperatures. The enrichment regime extends further than drag because, when $0<D_\mathrm{pd}^\mathrm{vac}<1$, solutes diffusing by the inverse Kirkendall mechanism are slower than host atoms, especially for Ni where $D_\mathrm{pd}^\mathrm{vac}<1$ up to much higher temperatures than other solutes (Mn, Cu, Si) with similar or stronger drag tendencies. On the other hand, Cr depletion occurs because vacancy drag is absent above 260 K and Cr diffusion is faster than self diffusion.

The dumbbell PDC ratios show a clear demarcation between solutes undergoing enrichment (P, Mn, Cr) and depletion (Cu, Ni, and Si up to 1310 K), which only incidentally corresponds to the categorization in stable and unstable mixed dumbbells. For P and Mn, dumbbell-induced enrichment is expected due to the MD stability and mobility discussed in Section \ref{sec:transport_clusters}. The difference between Cr and Si, on the other hand, is in apparent contradiction with the small difference between the respective MD binding energies, but can be explained by the fact that the $\omega_1$ migration barrier is lower than $\omega_0$ for Cr (0.24 vs 0.33 eV), while higher than $\omega_0$ for Si (0.57 vs 0.33 eV). 

\section{Radiation-induced segregation}

The transport coefficients  are here combined to analyze the intrinsic solute RIS tendencies stemming from the coupling with PDs. To this extent, the steady-state solution of Wiedersich's model \cite{wiedersich_theory_1979} is used. This was later adapted by Nastar, Soisson, and Martinez \cite{nastar_radiation-induced_2012, piochaud_atomic-based_2016, martinez_role_2018} to define the partial and intrinsic diffusion coefficients in terms of transport coefficients and alloy/PD driving forces. In this model, the relationship between vacancy and solute concentration gradients can be written as:
\begin{equation}
\nabla C_\mathrm{B} = - \alpha \frac{\nabla C_\mathrm{V}}{C_\mathrm{V}} \; ,
\end{equation}
where factor $\alpha$ is:
\begin{gather}
\alpha = \frac{\ell_\mathrm{AI}\ell_\mathrm{AV} } {\ell_\mathrm{AI}\left( D_\mathrm{B}^\mathrm{vac} + K D_\mathrm{B}^\mathrm{sia} \right) + \ell_\mathrm{BI}\left( D_\mathrm{A}^\mathrm{vac} + K D_\mathrm{A}^\mathrm{sia} \right) } \left[ \frac{ C_\mathrm{B} \left( D_\mathrm{pd}^\mathrm{vac} - D_\mathrm{pd}^\mathrm{sia} \right) }{ 1-C_\mathrm{B} }  \right] \; ,
\label{eq:alfa_ris}
\end{gather}
with $\ell_{i\delta}=L_{i\delta}/C_\delta$ ($\delta = \mathrm{V, I}$). The definition of the normalized intrinsic diffusion coefficients ($D_\mathrm{A}^\mathrm{vac}$,  $D_\mathrm{B}^\mathrm{vac}$, $D_\mathrm{A}^\mathrm{sia}$, $D_\mathrm{B}^\mathrm{sia}$), more details about the underlying assumptions, and the analytical derivations can be found in \ref{sec:appendix_ris}, in the formula summary of Table \ref{tab:formula_summary}, and in the original works \cite{wiedersich_theory_1979, nastar_radiation-induced_2012, martinez_role_2018}. Since the vacancy gradient is negative near the sink interface, the sign of $\alpha$, controlled by the difference of PDC ratios (Eq. \ref{eq:pdc_ratio}), determines whether solute enrichment ($\alpha > 0$) or depletion ($\alpha < 0$) occurs. 

In Eq. \ref{eq:alfa_ris}, $\alpha$ is in first approximation proportional to  $C_\mathrm{B}$. Factor $K=C_\mathrm{I}/C_\mathrm{V}$ groups all dependencies on $C_\mathrm{V}$ and $C_\mathrm{I}$, and can be adapted to specific irradiation and microstructure conditions. In order to analyze RIS regardless of any external parameter other than the intrinsic flux-coupling tendencies, it is chosen to set $C_\mathrm{I}/C_\mathrm{V} = D_\mathrm{V}\mathrm{(Fe)}/D_\mathrm{I}\mathrm{(Fe)}$, where $D_\mathrm{V}\mathrm{(Fe)}$ and $D_\mathrm{I}\mathrm{(Fe)}$ are the diffusion coefficients of PDs in pure Fe (i.e., solute enhancement is neglected). This choice of $C_\mathrm{I}/C_\mathrm{V}$ corresponds to the steady-state solution of the rate-theory equations describing the evolution of the PD population, when PD recombination, as well as sink and source bias, are neglected  \cite{sizmann_effect_1978}.

\begin{figure}[!htb]
	\centering
	\resizebox{0.6\columnwidth}{!}{
		\includegraphics{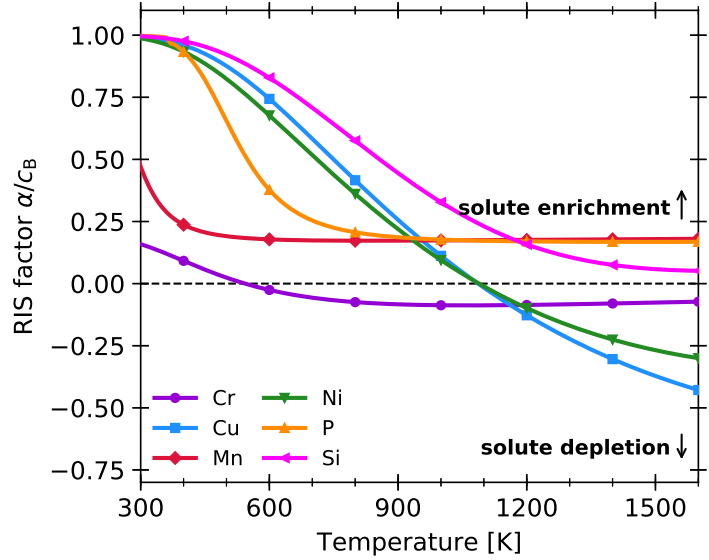}}
	\caption{\label{fig:ris} Radiation-induced segregation (RIS) tendencies (factor $\alpha$ in Eq. \ref{eq:alfa_ris} divided by the solute concentration $C_\mathrm{B}$) obtained with the KineCluE transport coefficients at $C_\mathrm{B}=0.1\%$. Solute enrichment occurs for positive $\alpha$ values.}
\end{figure}

Figure \ref{fig:ris} shows the $\alpha/C_\mathrm{B}$ ratio obtained for $C_\mathrm{B}=0.1\%$; this ratio is roughly independent of $C_\mathrm{B}$ except a small effect on the normalized intrinsic diffusion coefficients. It is worth reminding that non-linear solute concentration effects are not included in this dilute model, due to the lack of multiple-solute and multiple-defect interactions. Relevant temperatures for vacancy drag and RIS, alongside qualitative indications of the dominant mechanisms, are reported in Table \ref{tab:results_recap}. 

The model predicts systematic enrichment of P, Mn, and Si, and enrichment below $\approx 1085$ K for Cu and Ni, with mutually similar behaviors. In FeCr alloys, a switchover between Cr enrichment and depletion takes place at temperatures relevant for nuclear structural materials ($543 \pm 30$ K, depending on the chosen, yet dilute, Cr concentration). Based on the vacancy and dumbbells PDC  ratios shown in Fig. \ref{fig:pdc_ratios}, it is possible to identify the dominant diffusion mechanism determining the RIS behavior of each solute. 
\begin{itemize}
	\item[-] In FeCu, the depletion contribution of dumbbells is negligible. The total RIS behavior is thus controlled by vacancy drag: the maximum drag temperature coincides with the switchover between enrichment and depletion (1086 K). 
	\item[-] The behavior of Ni is apparently identical to Cu, with a switchover at the same temperature. Indeed, in the vacancy-drag regime Ni enrichment takes place. However, outside the drag regime the vacancy and dumbbell PDC ratios are close to one another, and both comprised between 0 and 1. So, even though vacancy diffusion would still yield an enrichment tendency, depletion eventually occurs because of the dumbbell contribution. The RIS tendency is therefore controlled by vacancies below 1089 K, and by dumbbells above.  
	\item[-] Si follows a similar reasoning as Ni, but with a different conclusion. Vacancy drag determines Si enrichment up to the temperature at which it disappears (1308 K). Above that, dumbbell diffusion provides an enrichment tendency that overturns vacancy-driven depletion, keeping Si in an enrichment regime in spite of the absence of vacancy drag.
	\item[-] In FeP and FeMn, both mechanisms yield enrichment, but the dumbbell contribution is dominant by several orders of magnitude. Hence, P and Mn enrichment occurs almost exclusively by dumbbell transport.
	\item[-] For Cr, the two mechanisms give rise to opposite tendencies with similar magnitudes, as pointed out in previous studies \cite{wharry_mechanism_2014, senninger_modeling_2016}. The bottom panel of Fig. \ref{fig:senninger_comparison} (thick red curve) shows that the total PDC ratio determining the sign of $\alpha$ remains very close zero in a wide range above the switchover temperature ($T_\mathrm{RIS}$). Any minor influence from various parameters or external conditions can thus lead to large changes in $T_\mathrm{RIS}$. This is discussed in more detail in Section \ref{sec:discussion_senninger}.  
\end{itemize} 

Quite surprisingly, the RIS magnitude is smaller in a dumbbell-driven regime than in a vacancy-driven one. This is the case for P and Mn, as well as for Si in the temperature range where dumbbells are dominant. This might be explained in terms of back diffusion, i.e., defect fluxes attempting to restore a homogeneous concentration gradient near sinks. The efficiency of solute-dumbbell transport might be favoring RIS and contrasting it at the same time by facilitating back diffusion, thus reducing the total effect.

\section{Discussion}

\subsection{Solute-dumbbell transport in view of previous \textit{ab initio} and experimental studies}
\label{sec:discussion_DFT}

The obtained dumbbell transport properties are discussed here in terms of previous \abinitio calculations, evidence from RR experiments, and common empirical assumptions.

In the FeP system, P migrates in a continuous oscillation between the MD and the '1b' configuration, with a very small dissociation probability. The combination of strong binding and high mobility produces a very long-ranged migration that goes far beyond common grain sizes (cf. Fig. \ref{fig:mfp}). These results confirm previous RR experiments according to which the Fe-P dumbbell has a higher mobility than self-interstitials, and slightly higher than Fe-Cr dumbbells \cite{abe_interaction_1999}. This seems to exclude the presence of trapping configurations. However, previous \abinitio works \cite{domain_diffusion_2005, meslin_theoretical_2007} have shown that the addition of a second P atom has a strong trapping effect on the Fe-P dumbbell. Therefore, P transport might become less effective with growing solute concentration. On the other hand, this may be counterbalanced by the fact that P migration is likely to involve foreign interstitial sites. According to previous {\sc siesta} calculations \cite{meslin_theoretical_2007}, the P formation energy in octahedral sites is lower than in the MD configuration, and a MD jump passing through the octahedral configuration has actually a lower barrier (0.17 eV) than the RT mechanism. In addition, the barrier for a 2nn MD jump was found to be as low (0.18 eV). In this work, the difference between MD and octahedral stability (+0.25 eV) is actually larger than previous calculations with {\sc siesta} (-0.08 eV \cite{meslin_theoretical_2007}) and {\sc vasp} ultrasoft pseudopotentials (+0.05 eV \cite{domain_diffusion_2005}), so the octahedral diffusion pathway might be less important. Nevertheless, the matter should be further investigated. 

The Fe-Mn dumbbell is very stable and has a mobility close to that of self-interstitials, as was suggested by RR experiments \cite{maury_interstitial_1990} and later proven by AKMC simulations of isochronal annealing \cite{ngayam-happy_isochronal_2010}. The resulting MFPs are not as long as for Fe-P, but still comparable to common grain sizes. In a regime of defect absorption at sinks, hence, P and Mn can be expected to reach grain boundaries rather easily. For this reason, the frequent observation of P/Mn grain-boundary segregation is unsurprising. In addition, previous \abinitio calculations have found a very low MD association barrier via a 2nn jump (0.04 eV) \cite{vincent_ab_2006}. If confirmed, this would entail that the solute-dumbbell pair dissociation is practically impossible, and may increase correlations reducing the diffusivity of Mn solute atoms.  

The Fe-Cr mixed-dumbbell interaction is attractive but weak, similarly to the vacancy-Cr interaction \cite{messina_exact_2014}. 
Moreover, the 'M' and '1b' binding energies are very close to one another. According to earlier DFT calculations based on the Perdew and Wang (PW91) functional, the MD should be slightly more stable \cite{olsson_ab_2010}. PBE is in agreement with PW91 only when the accuracy is increased to $4\times 4\times 4$ k-points and 400 eV. The obtained migration barriers are in good agreement with a previous work \cite{choudhury_ab-initio_2011}, except a mismatch in the MD association/dissociation rate.  The high MD mobility is in accordance with RR  experiments \cite{maury_study_1987} and previous molecular-dynamics simulations \cite{terentyev_migration_2008}. However, the MFP of the Fe-Cr dumbbell is rather short with respect to Fe-P and Fe-Mn. This means that Cr migration occurs by exchange with several Fe-Fe dumbbells that in turn form a mixed dumbbell, displace the Cr solute by a few {\aa}ngstr\"om, and then quickly dissociate. It should also be mentioned that a 2nn association jump is possible with a barrier of 0.36 eV, as pointed out by Olsson \cite{olsson_ab_2009}.

Si and Ni present the most surprising behavior. Whereas RR experiments were interpreted by assuming the formation and migration of the Fe-Si and Fe-Ni mixed dumbbells \cite{maury_interstitial_1985, abe_interaction_1999}, there is no unanimous consensus on the effective possibility of Si/Ni migration by a dumbbell mechanism. A previous \abinitio study excluded this hypothesis for both solutes based on the unfavorable MD repulsive energy and the high dissociation probability \cite{vincent_ab_2006}. Indeed, it is confirmed here that the Fe-Si and Fe-Ni are not stable, have a strong tendency to dissociate (low $\omega_{\mathrm{M}\,\mathrm{1b}}$ barrier), and their MFP before dissociation is essentially null. However, since for both Si and Ni the vacancy- and dumbbell-related diffusion coefficients (divided by defect concentration) have comparable magnitudes (cf. Fig. \ref{fig:solute_diff}), 
the dominant mechanism is determined by the ratio $C_\mathrm{V}/C_\mathrm{I}$. When the latter is larger than 1, vacancy diffusion dominates: for instance, if $C_\mathrm{V}/C_\mathrm{I} \approx D_\mathrm{I}/D_\mathrm{V}$, this ratio is roughly equal to $10^3$ at the RPV operation temperature of roughly 300 \degree{}C, which indicates that under these assumptions Si and Ni diffuse predominantly by vacancies. However, this can change depending on the specific microstructure and irradiation conditions. It is also worth mentioning that the same solutes were observed to trap self-interstitial atoms in RR experiments \cite{maury_interstitial_1985, maury_interstitial_1990}. In FeSi, this can be clearly ascribed to the '1b' configuration (strong binding and high migration barriers). In FeNi, it could be due to the '1b' configuration as well, although with a weaker binding strength. 

Finally, the strongly repulsive interaction of the FeCu mixed dumbbell, combined with its spontaneous dissociation (zero $\omega_{\mathrm{M}\,\mathrm{1b}}$ barrier), confirms the common assumption that Cu diffusivity by dumbbells is negligible. However, Cu can affect the migration of self interstitials by trapping them in the '1a' configuration: the migration barriers towards '1a' are sensibly lower than the reverse ones, and also lower than $\omega_0$. This trapping configuration was shown in isochronal annealing simulations \cite{ngayam-happy_isochronal_2010} to be the cause for the observed disappearance of the single-SIA migration peak in RR experiments \cite{maury_interstitial_1990}. The same experiment mentions as well further trapping caused by a second Cu atom, so the slowdown effect on dumbbells might grow with Cu concentration.

\subsection{Prediction of dumbbell transport and RIS for other Fe-based alloys}
Earlier systematic investigations 
allow for a qualitative prediction of solute transport and RIS for the other transition-metal solutes in Fe. The vacancy-transport properties have been accurately determined with an analogous \textit{ab initio}-SCMF model \cite{messina_systematic_2016}. After removing the vacancy formation energy, the solute diffusion activation energies range between 0.2 and 0.9 eV, with a bell-shaped profile across the 4d and 5d elements reaching a maximum in the mid-row elements (Ru and Os). Vacancy drag and solute enrichment extends to high temperatures for the early and late elements. For the mid-row elements (and Ti), the maximum drag temperature is in the range 500-700 K, and depletion occurs above this range. A few exceptions were mentioned: depletion due to the lack of vacancy drag takes place for V and Cr, while for a set of "slow diffusers" (Co, Re, Os, Ir) enrichment always occurs because of the lower diffusivity than host atoms.

There has been no systematic analysis of the same kind for dumbbell diffusion, but a systematic calculation of MD binding energies is available \cite{olsson_ab_2010}. All MDs but Fe-Mn and Fe-Cr have a strong repulsive interaction ($<-0.8$ eV), except Co and V that have a milder repulsion ($-0.25$ and $-0.55$ eV, respectively). Based on this evidence, it would be tempting to assume that none of these solute diffuse by dumbbells. However, this must be carefully assessed by comparing the activation energies. For dumbbell diffusion, it holds $E_\mathrm{act}^\mathrm{I} \approx E_\mathrm{mig}^\mathrm{I} - E_\mathrm{bind}^\mathrm{I}$. Hence, neglecting the prefactors, solute diffusion by dumbbells is quicker than by vacancies if $E_\mathrm{act}^\mathrm{I} < E_\mathrm{act}^\mathrm{V}$, i.e., if $E_\mathrm{mig}^\mathrm{I} < E_\mathrm{bind}^\mathrm{I} + E_\mathrm{act}^\mathrm{V}$. Since the MD migration energies $E_\mathrm{mig}^\mathrm{I}$ cannot be negative, combining the vacancy activation energies with the MD binding energies leaves only two candidates: vanadium, if the MD migration energy is lower than 0.19 eV, and cobalt,  if lower than 0.58 eV. The latter case is definitely possible, given that for instance the Fe-Cu MD has a migration barrier of 0.36 eV in spite of a strong repulsive interaction (-0.38 eV).   

By a similar logic, it is possible to make predictions on the dumbbell PDC ratios. In this case, if flux coupling is negligible (which is not necessarily the case), the PDC ratio can be evaluated by comparing the MD activation energy with that of the Fe-Fe dumbbell: namely, $D_\mathrm{pd}^\mathrm{I}>1$ if $E_\mathrm{mig}^\mathrm{I} < E_\mathrm{bind}^\mathrm{I} + E_\mathrm{mig}^0$, where $E_\mathrm{mig}^0$ is the Fe-Fe  migration energy (0.33 eV). The only solute that might fulfill this condition is cobalt, if the Fe-Co migration energy were smaller than 0.08 eV, which seems unlikely. Thus, it is possible to conclude that depletion by dumbbells is expected for all solutes ($D_\mathrm{pd}^\mathrm{I}<1$). It is therefore expected to have vacancy-induced enrichment in the vacancy-drag regime, and vacancy-induced depletion in the absence of drag when $D_\mathrm{pd}^\mathrm{V}>1$. In the interval $0 < D_\mathrm{pd}^\mathrm{V}<1$, there might be a competition between the two mechanisms, depending on the magnitude of  $D_\mathrm{pd}^\mathrm{I}$. For slow diffusers, and especially Co, this prediction  must be supported by further \abinitio calculations, including at least the MD dumbbell migration energy. 

\subsection{Comparison with previous transport coefficient calculations}
\label{sec:discussion_senninger}

Full sets of Onsager coefficients have been previously computed only for the FeCr and FeNi systems. Chodhury \textit{et al.} \cite{choudhury_ab-initio_2011} used the Le Claire/Serruys model \cite{le_claire_solute_1978, serruys_determination_1982} for vacancies and Barbe's SCMF model \cite{barbe_phenomenological_2007} for dumbbells to compute the full matrix, but then based the RIS discussion on the ratio of diffusion coefficients $D_\mathrm{B}/D_\mathrm{A}$, without taking flux coupling into account. The PDC ratios in Fig. \ref{fig:pdc_ratios} compare qualitatively well with Choudhury's results concerning Cr-dumbbell, Ni-dumbbell, and Cr-vacancy diffusion, since flux coupling in these systems is negligible. On the contrary, for Ni-vacancy diffusion the ratio is smaller than 1, while according to Choudhury's model it is consistently larger than 1 (i.e., depletion) due to the lack of flux coupling.

\begin{figure}[!htb]
	\centering
	\resizebox{0.6\columnwidth}{!}{
		\includegraphics{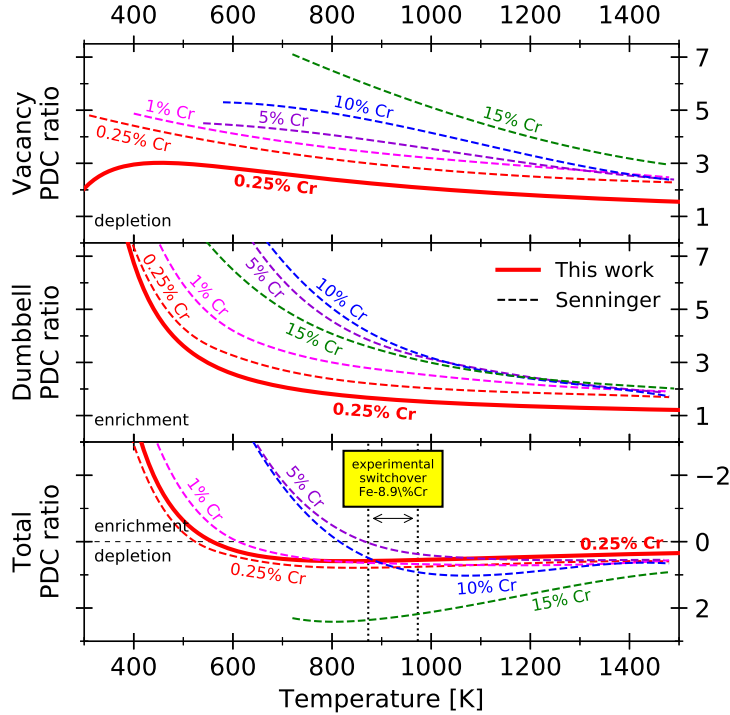}}
	\caption{\label{fig:senninger_comparison} Comparison between the partial diffusion coefficient (PDC) ratios obtained in this work in the dilute limit, and the Cr concentration-dependent ones computed by Senninger \textit{et al.} \cite{senninger_modeling_2016} with an AKMC model. From top to bottom: PDC ratios for vacancy diffusion ($D_\mathrm{pd}^\mathrm{V}$), for dumbbell diffusion ($D_\mathrm{pd}^\mathrm{I}$), and difference ($D_\mathrm{pd}^\mathrm{V}-D_\mathrm{pd}^\mathrm{I}$) determining the sign of the global RIS tendency (enrichment when negative). The experimental switchover range refers to the RIS measurements by Wharry \textit{et al.} in a T91 alloy \cite{wharry_systematic_2013}.  }
\end{figure}

A series of combined AKMC-phase field studies, based on an \textit{ab initio}-fitted pair interaction model, has achieved a more accurate analysis of Cr segregation in FeCr, including the effects of flux coupling and non-dilute concentrations \cite{senninger_modeling_2016, piochaud_atomic-based_2016, thuinet_multiscale_2018}. The AKMC-computed PDC ratios are compared to the results of this work in Fig. \ref{fig:senninger_comparison}, where the bottom panel reports the difference between vacancy and dumbbell PDC ratios. Our dilute model does not provide the variation with Cr concentration, but in the most dilute case (0.25\% Cr) it compares very well with Senninger's results \cite{senninger_modeling_2016} both in terms of switchover temperature $T_\mathrm{RIS}$ and magnitude. This clearly proves that the mismatch in $T_\mathrm{RIS}$ between KineCluE and the experimental observation is at least partially due to Cr concentration effects. It is also possible to observe that, according to Senninger's work, $T_\mathrm{RIS}$ has a maximum somewhere between 5\% and 10\% Cr, before decreasing again, in striking agreement with Wharry's experimental observations \cite{wharry_systematic_2013}. This occurs because, while vacancy depletion increases monotonically, dumbbell enrichment reaches a maximum at 10\% Cr and then decreases. The mobility of the Fe-Cr dumbbell seems thus to be reduced by the presence of more Cr atoms, in accordance with Terentyev's molecular-dynamics simulations \cite{terentyev_migration_2008}.

Vacancy depletion and dumbbell enrichment in FeCr have very similar magnitudes. This makes the RIS behavior very sensitive to any parameter that can perturb this balance, and is most likely the reason why conflicting conclusions (enrichment vs depletion) can be found in the literature for ferritic alloys \cite{takahashi_radiation-induced_1981, ohnuki_void_1981, kato_effect_1991}. Besides Cr concentrations, other parameters such as sink density, sink and source bias, and strains may affect RIS. In particular, strain fields induced by defect sinks are surely playing a major role. Thuinet \textit{et al.} \cite{thuinet_multiscale_2018} performed a first strain-dependent investigation of RIS in FeCr alloys with phase-field simulations, and concluded that vacancy-attracting compression zones should be more prone to Cr depletion (the opposite for traction zones). This is sufficient to overturn the results of the strain-free model. However, these conclusions were drawn neglecting the effect of strains on flux coupling. Since strain-dependent transport coefficients can be computed in KineCluE, this topic will be investigated in future studies.

\subsection{Experimental diffusion coefficients and finite-temperature effects}

Experimental data for dumbbell-assisted diffusion are very difficult to obtain and are therefore not available for comparison. For vacancy migration, the solute diffusion coefficients computed in the SCMF framework were already successfully compared to experimental measurements in previous publications \cite{messina_exact_2014, messina_systematic_2016}. A very close match in activation energies and only slight differences in the prefactors (one order of magnitude or less) were found. In these earlier works, a magnetic model \cite{sandberg_modeling_2015} accounted for the effect of the magnetic transition, in the aim of a meaningful comparison with experiments in the range around the Curie temperature. In that model, it was assumed that the variation of the magnetic enthalpy with temperature had an absolute effect on the self-diffusion activation energy, but no relative effects on solute behavior. This assumption proved valid for all solutes except Mn, which is known to hold odd magnetic properties in Fe that are currently under investigation \cite{schneider_local_2018}. For these reasons, no magnetic model is implemented in this work. Conclusions on solute-transport properties and RIS are still valid, especially at low temperatures  where finite-temperature effects due to magnetism are negligible \cite{sandberg_modeling_2015}. Vibrational-entropy harmonic contributions to the defect formation and solute migration energies are included. However, this is not the case for the solute-defect binding entropy, which can lead to variations of the relative stability of solute-defect pairs, as was shown for instance by Murali \textit{et al.} \cite{murali_first-principles_2015}.

\subsection{Comparison with experimental RIS observations}

The RIS tendencies presented in Fig. \ref{fig:ris} are consistent with many experimental observations. Firstly, as pointed out by Rehn and Okamoto, no case of depletion of undersized solute species in dilute alloys has ever been reported \cite{okamoto_radiation-induced_1979}. According to Ardell \cite{ardell_radiation-induced_2016}, this holds true still today. The consistent enrichment of the undersized elements (P and Si) shown in Fig. \ref{fig:ris} confirm these observations.

Furthermore, our results match well those of Wharry's RIS study in several FeCr alloys between 300 and 700 \degree{}C \cite{wharry_systematic_2013} . Wharry found a systematic enrichment of Si, Cu, and Ni at grain boundaries. In addition, Cr was found to switch from (weak) enrichment to depletion across 600-700 \degree{}C. In view of this difference, they suggested that the mechanisms driving the RIS behavior of these solutes might be different. Indeed, the results presented here confirm all of these findings, and prove that Si/Cu/Ni enrichment is due to vacancy drag, while Cr enrichment to dumbbell transport. The mismatch in switchover temperature (543 K vs $\approx$ 900 K) can be ascribed to its dependence on solute concentration, as shown in Fig. \ref{fig:senninger_comparison} and discussed in Section \ref{sec:discussion_senninger}.

In support to our findings, P and Mn have been repeatedly observed to segregate on grain boundaries, dislocation lines, and loops \cite{miller_atom_2013, meslin_radiation-induced_2013, pareige_behaviour_2015}. Furthermore, all solutes considered here have been observed to form vacancy-solute clusters in model alloys and RPV steels \cite{nagai_positron_2003, glade_positron_2006, konstantinovic_thermal_2015}, which is consistent with a positive coupling between solutes and point defects. In RPV steels, the thermal stability of Mn-Ni-Si clusters \cite{sprouster_structural_2016, ke_thermodynamic_2017, almirall_elevated_2019, almirall_mechanistic_2020} points towards the existence of a thermodynamic driving force for precipitation, although the actual solubility limit in these complex multi-component alloys is yet to be precisely determined. In this context, positive solute RIS tendencies should enhance the heterogeneous nucleation of Mn-Ni-Si-rich secondary phases at PD sinks, such as pinned SIA clusters \cite{bonny_monte_2014, castin_dominating_2020}. Our results show that Mn, Ni, and Si can indeed be dragged to sinks by point defects, and thus confirm the important role of kinetic coupling in the precipitation process.

Similar phenomena have been observed in other alloys. For instance, Ni enrichment at dislocation loops has been observed to occur in FeNi alloys, as Ni atoms bind with loops and then act as sinks for mobile defect-Ni complexes. This mechanism could be the precursor of the observed secondary-phase precipitation \cite{belkacemi_radiation-induced_2018}. Our results suggest that such mobile defect-Ni complexes are likely to be vacancy-Ni pairs. In Cr-rich ferritic alloys, Cr-Si-Ni-P clusters are formed due to P atoms diffusing to dislocations and catalyzing the formation and growth of clusters around them  \cite{pareige_behaviour_2015, gomez-ferrer_role_2019}. Fig. \ref{fig:solute_diff} shows that P is indeed the fastest diffuser thanks to the Fe-P dumbbell, which might thus be responsible for the onset of cluster precipitation in these alloys.

\begin{figure}[!htb]
	\centering
	\resizebox{0.6\columnwidth}{!}{
		\includegraphics{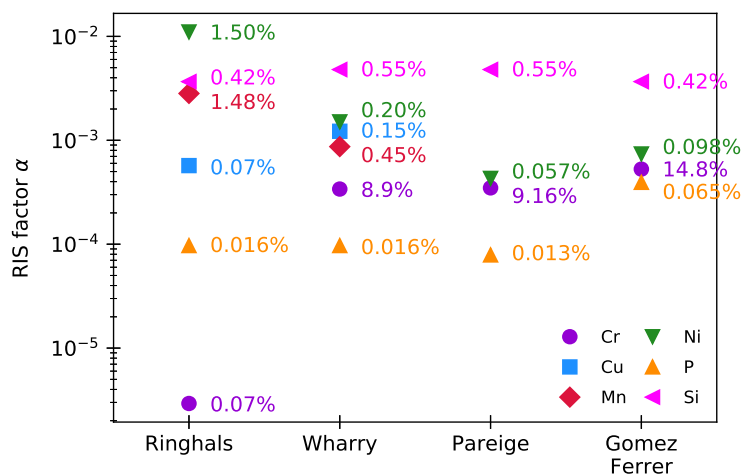}}
	\caption{\label{fig:ris_alloys} KineCluE radiation-induced segregation (RIS) factors $\alpha$ (cf. Eq. \ref{eq:alfa_ris}) marking solute enrichment in several multicomponent alloys, namely an RPV steel (Ringhals) \cite{efsing_long_2013}, and three FeCr alloys: T91 (Wharry) \cite{wharry_systematic_2013}, Fe-9\%Cr (Pareige) \cite{pareige_behaviour_2015}, and Fe-15\%Cr (Gomez-Ferrer) \cite{gomez-ferrer_role_2019}. Each data point reports also the nominal concentration (in at.\%) of the corresponding element.} 
\end{figure}

Figure \ref{fig:ris_alloys} reports a more quantitative evaluation of solute segregation in the following alloys: the Ringhals high-Ni, low-Cu RPV steel \cite{efsing_long_2013} and three FeCr alloys, including Wharry's T91 \cite{wharry_systematic_2013} and the two model alloys where P-catalyzed precipitation took place \cite{pareige_behaviour_2015, gomez-ferrer_role_2019}. The solute nominal concentrations are reported next to each data point. The calculation of the transport coefficients has been modified in order to add a "multicomponent effect" to the total partition function (cf. \ref{sec:appendix_multicomponent} for the mathematical details). This is to take into account the fact that, in a multicomponent system with fixed PD concentration, the mobility of a given solute might be reduced if the majority of PDs binds preferentially to other solutes. For instance, in the Ringhals RPV steel the vast majority of SIAs is bound to P atoms, and just a small fraction ($\approx 0.6\%$ of the total dumbbell population at 300 \degree{}C according to our calculations) is bound to Mn atoms, even though the nominal P concentration is much lower (0.016\% vs 1.48\%) and the formation of Fe-Mn dumbbells is favorable. Consequently, Mn diffusivity in the multicomponent alloy is lower than in a binary FeMn alloy, which might explain why Ni segregation (predominantly driven by vacancies) is higher than Mn segregation even though the starting Mn and Ni concentrations are similar. However, the predicted segregation magnitudes are to some extent in disagreement with the experimental compositions of Mn-Ni-Si clusters, showing a majority of Mn and Ni atoms, with $\mathrm{Ni}>\mathrm{Mn}$, a lower Si content, and traces of Cu and P \cite{miller_atom_2013}. Such mismatch in cluster composition likely supports the argument that RIS is the precursor and an important enhancement factor of the formation of some stable phases, but not the only cause. Concerning the FeCr alloys, the segregation magnitudes compare well with Wharry's study \cite{wharry_systematic_2013}, in which Si, Ni and Cu segregation (in decreasing order of magnitude) was observed, but less well in the other cases: our model yields a stronger Si segregation than Ni, in place of the experimentally observed similar magnitudes. 

Let us summarize the limitations of our RIS model. The latter provides the kinetic contribution to segregation due to solute-PD coupling in a dilute limit approach, i.e., it neglects non-linear solute-concentration effects and the possible coupling with small PD clusters, which might be relevant when solutes and PDs accumulate locally next to sinks. Under such conditions, the solute driving forces, not considered in this model, might reduce solute back fluxes and promote segregation due to solute stabilization. Furthermore, any effect of sink absorption bias, local sink-induced strains, and binding free-energy variations due to finite temperatures is neglected, and might affect the predicted temperature thresholds. More importantly, the thermodynamic interaction between solutes and sinks is currently not included in the model, and might add a localized equilibrium segregation profile in addition or in contrast to RIS. Typical examples of such interplay are the so-called "W-shaped" profiles arising when RIS promotes solute depletion at sinks characterized by attractive thermodynamic interactions with solutes \cite{nastar_radiation-induced_2012}.
	
Introducing chemical potential gradients, in combination with specific irradiation and microstructure conditions, would allow for a more significant comparison to experiments, and for the prediction of the bell-shaped segregation profiles found, for instance, by Wharry \textit{et al.} \cite{wharry_systematic_2013}. This shape arises because, at low temperature, segregation is limited by slow diffusivities, whereas at high temperature, the total PD driving force gets progressively weaker due to the increasing equilibrium concentrations. It must be mentioned, however, that according to Mart\`inez \textit{et al.} \cite{martinez_role_2018}, for low equilibrium PD concentrations the total amount of segregated solutes depends only on the RIS factor $\alpha$ and the sink density, even when the PD driving forces are taken into account. For this reason, the comparison shown in Fig. \ref{fig:ris} is still relevant, and the fundamental solute transport behaviors here uncovered can be used as a guide for experimental interpretations and microstructure models. 

\section{Conclusions}

This work has provided an in-depth investigation of solute diffusion by interstitial-type defects in binary dilute ferritic alloys including Cr, Cu, Mn, Ni, P, and Si solute atoms. In combination with previous knowledge on vacancy-assisted diffusion, a general overview of the intrinsic radiation-induced segregation (RIS) behavior of these solutes has been achieved. The most relevant findings can be summarized as follows.
\begin{enumerate}
	\item P, Mn, and Cr form stable mixed dumbbells, while the Fe-Si dumbbell is neither binding nor repulsive. The Fe-Cu and Fe-Ni dumbbells are repulsive at 0 K, but due to soft vibrational modes, they might be more stable at higher temperatures. Si, Cu, and Ni solutes can trap Fe-Fe dumbbells in a first-nearest neighbor position.
	\item Vacancy-solute pairs are stable and mobile. At 300 \degree{}C, their mean free paths can reach up to 1 nm, and up to 10 nm for P. The Fe-Cu, Fe-Ni, and Fe-Si dumbbells have a negligible mobility and a high dissociation rate, while the Fe-Cr, Fe-P, and Fe-Mn dumbbells are mobile. The mean free path of a single Fe-P and Fe-Mn dumbbell can exceed 1 \textmu m. 
	\item Even if the mixed dumbbell is unstable, dumbbell diffusion of solutes can be in competition with vacancy diffusion. Ni diffusivity by vacancies (normalized to the vacancy concentration) is five times faster than dumbbells at 300 \degree{}C and only twice as fast at 630 \degree{}C. For Si atoms, vacancy diffusion is slightly predominant below 300 \degree{}C, but perfectly equivalent to dumbbell diffusion at RPV temperatures. Dumbbells provide the predominant contribution to P, Mn, and Cr diffusivity, while vacancies dominate that of Cu. P is the fastest diffusing species by both mechanisms. The predominance of a mechanism over the other depends on the ratio of point defect concentrations $C_\mathrm{V}/C_\mathrm{I}$. Among all other transition-metal (TM) solutes, Co is the only one for which the dumbbell mechanism might be comparable with the vacancy one.
	\item Systematic enrichment is predicted for P and Mn due to dumbbell diffusion, and for Si due to vacancy drag. Cu and Ni enrichment by vacancy drag takes place below 1085 K. Right above this limit, depletion should occur. Enrichment by vacancy drag is dominant for all other TM solutes when present \cite{messina_systematic_2016}; in the opposite case, dumbbell depletion can be in competition with inverse-Kirkendall diffusion.  
	\item The RIS behavior of Cr is the outcome of a fine balance between dumbbell enrichment and vacancy depletion, and this leads to a switch from enrichment to depletion at $T_\mathrm{RIS}=540$ K. The match is so close that small variations of any parameter (e.g., Cr concentration, strains, sink biases) can lead to large changes of $T_\mathrm{RIS}$. This is in agreement with experimental RIS observation and previous calculations, which pointed out that $T_\mathrm{RIS}$ grows with Cr concentration up to about 10\% Cr, where $T_\mathrm{RIS}$ is comprised between 600 and 700 \degree{}C. 
	\item These findings are consistent with the interpretation of the known RIS and solute clustering phenomena. They confirm that a drag mechanism by point defects is active for all solutes, and can provide an important kinetic mechanism that enhances the formation of solute clusters and precipitates in many types of Fe alloys, including RPV alloys. 
\end{enumerate}

This study was the first application of a novel, multi-scale framework that combines \abinitio calculations (or any other energy model) with the self-consistent mean-field model and a computationally efficient numerical tool (KineCluE) to produce the transport coefficients of an alloy to a high degree of accuracy. This can be used to analyze solute transport and diffusion properties, flux coupling with point defects, and RIS behaviors for a wide range of periodic crystal structures and diffusion mechanisms. Even though the current study is limited to kinetic segregation in the dilute-limit approach, it can be extended by adding thermodynamic interactions with sinks and the energetics of clusters larger than single defect-solute pairs, with the aid of an appropriate energy model. It also allows for strain-dependent calculations of transport coefficients, which will be the object of future studies. The presented framework is a valuable asset for gaining fundamental knowledge that can be useful to interpret experimental observation and devise physically accurate microstructure evolution models.

\vspace{0.8cm}

\section*{Acknowledgments}
This work was financially supported by Vattenfall AB, G\"{o}ran Gustafsson Stiftelse, and the Euratom research and training programme 2014-2018 under the grant agreement No 661913 (SOTERIA). The paper reflects only the authors' view and the European Commission is not responsible for any use that may be made of the information it contains. It contributes to the Joint Program on Nuclear Materials (JPNM) of the European Energy Research Alliance (EERA). The high-performance computing resources were provided by the Swedish National Infrastructure for Computing (SNIC) and the GENCI (CINES/CCRT) computer center under Grant No. A0070906973. Moreover, the authors acknowledge T. Garnier for the valuable contributions to the SCMF method.  

\section*{References}
\bibliography{manuscript_arXiv_zotero.bib}

\appendix

\section{Configurations and jump frequencies for dumbbell-solute pairs in bcc}
\label{sec:appendix_coordinates}

Table \ref{tab:dumbbell_coords} reports the atomic coordinates corresponding to the dumbbell-solute configurations shown in Fig. \ref{fig:configurations}. Furthermore, Table \ref{tab:allowed_jumps} summarizes all dumbbell migration events in bcc crystals where the dumbbell moves by a 1nn distance, depicted also in Fig. \ref{fig:jump_nomenc}. Pure translations ($\tau$) and onsite rotations (R) are shown in addition to the rotation-translation jumps.

\begin{table}[htb]\scriptsize
	\caption{\label{tab:dumbbell_coords} List of symmetry-unique configurations of a dumbbell-solute pair up to the 5nn (cf. Fig. \ref{fig:configurations}). A and B mark the two atoms of the dumbbell, located respectively in $[-\delta,-\delta,0]$ and $[\delta,\delta,0]$; 2$\delta$ is the distance between them. The dumbbell is in the center of the reference system and oriented along the $\langle 110 \rangle$ direction. Lengths are given in units of $a_0/2$.  }
	\renewcommand{\arraystretch}{1.4}
	\centering
	\begin{adjustbox}{width=0.6\columnwidth}	
		\begin{tabular}{C{2cm}C{1cm}C{2cm}C{2cm}}
			\hline
			\textbf{Solute site}	&	\textbf{Label}	& \textbf{Distance to A}	&	\textbf{Distance to B}		\\
			\hline
			$[\delta, \delta, 0]$	&	M		&	$2\delta$			&	0			\\
			\hline
			$[-1, 1, 1]$	&	1a		&	$\sqrt{2\delta^2 + 3}$			&	$\sqrt{2\delta^2 + 3}$			\\
			$[1, 1, 1]$ 	&	1b		&	$\sqrt{2\delta^2 + 4\delta + 3}$	&	$\sqrt{2\delta^2 - 4\delta + 3}$		\\
			\hline
			$[0, 0, 2]$	&	2a		&	$\sqrt{2\delta^2 + 4}$			&	$\sqrt{2\delta^2 + 4}$			\\
			$[2, 0, 0]$	&	2b		&	$\sqrt{2\delta^2 + 4\delta + 4}$	&	$\sqrt{2\delta^2 - 4\delta + 4}$		\\
			\hline
			$[-2, 2, 0]$	&	3a		&	$\sqrt{2\delta^2 + 8}$			&	$\sqrt{2\delta^2 + 8}$			\\
			$[2, 0, 2]$	&	3b		&	$\sqrt{2\delta^2 + 4\delta + 8}$	&	$\sqrt{2\delta^2 - 4\delta + 8}$		\\
			$[2, 2, 0]$	&	3c		&	$\sqrt{2\delta^2 + 8\delta + 8}$	&	$\sqrt{2\delta^2 - 8\delta + 8}$		\\
			\hline
			$[-1, 1, 3]$	&	4a		&	$\sqrt{2\delta^2 + 11}$			&	$\sqrt{2\delta^2 + 11}$			\\
			$[1, 1, 3]$	&	4b		&	$\sqrt{2\delta^2 + 4\delta + 11}$	&	$\sqrt{2\delta^2 - 4\delta + 11}$	\\
			$[3, 1, 1]$	&	4c		&	$\sqrt{2\delta^2 + 8\delta + 11}$	&	$\sqrt{2\delta^2 - 8\delta + 11}$	\\
			\hline
			$[-2, 2, 2]$	&	5a		&	$\sqrt{2\delta^2 + 12}$			&	$\sqrt{2\delta^2 + 12}$			\\
			$[2, 2, 2]$	&	5b		&	$\sqrt{2\delta^2 + 8\delta + 12}$	&	$\sqrt{2\delta^2 - 8\delta + 12}$	\\
			
			\hline
			
		\end{tabular}
	\end{adjustbox}
\end{table}

\begin{table}[htb]\scriptsize
	\caption{\label{tab:allowed_jumps} Dumbbell migration events in a bcc crystal ($\tau$ for pure translation, $\omega$ for rotation-translation, and R for onsite rotation). Mixed-dumbbell transitions involving solute migration are marked with '1'. 2nn jumps (e.g., 2b$\rightarrow$M), as well as 1nn jumps involving sites further than the 2nn, are not shown.}
	\renewcommand{\arraystretch}{1.4}
	\centering
	\begin{adjustbox}{width=0.7\columnwidth}	
		\begin{tabular}{C{0.8cm}|C{1cm}|C{0.6cm}|C{0.6cm}|C{0.6cm}|C{0.6cm}|C{0.6cm}|C{0.6cm}|C{0.6cm}|C{0.6cm}|C{0.6cm}}
			\hline
			&	M	&	1a	&	1b	&	2a	&	2b	&	3b	&	3c	&	4b	&	4c	&	5b	\\
			\hline
			M	&	$\omega$, $\tau$, R	&	&	$\omega$, $\tau$	\\
			\hline	
			1a	&	&	R	&	R	&	$\omega$	&	$\omega$, $\tau$	&	$\omega$, $\tau$	&	$\omega$	\\
			\hline
			1b	&	$\omega$, $\tau$	&	R	&	R	&	$\tau$	&	$\omega$	&	$\omega$	&	$\tau$	&	&	&	$\omega$, $\tau$	\\
			\hline
			2a	&	&	$\omega$	&	$\tau$	&	&	R	&	&	&	$\tau$	&	$\omega$	\\
			\hline
			2b	&	&	$\omega$, $\tau$	&	$\omega$	&	R	&	R	&	&	&	$\omega$	&	$\omega$, $\tau$	\\
			\hline
			3b	&	&	$\omega$, $\tau$	&	$\omega$	\\
			\hline
			3c	&	&	$\omega$	&	$\tau$	\\
			\hline
			4b	&	&	\phantom{phan}&	\phantom{phan}	&	$\tau$	&	$\omega$	\\
			\hline
			4c	&	&	&	&	$\omega$	&	$\omega$, $\tau$ \\
			\hline
			5b	&	&	&	$\omega$, $\tau$	\\
			\hline
		\end{tabular}
	\end{adjustbox}
\end{table}

\section{Computation of total transport coefficients from cluster contributions}
\label{sec:appendix_computation_total_lij}

In the cluster development framework \cite{schuler_transport_2016, schuler_kineclue:_2020} of SCMF \cite{nastar_mean_2005}, each total transport coefficient $L_{ij}$ is obtained as a weighed sum of cluster contributions:
\begin{equation}
L_{ij} = \mathcal{C} \left[ \sum_m f_m L_{ij}^{(m)} + \sum_c f_c L_{ij}^{(c)} \right]  \; ,
\label{eq:total_lij}
\end{equation}
where $\mathcal{C}$ is the total concentration and $f_m$ ($f_c$) the fraction of each monomer $m$ (cluster $c$). The monomer and cluster transport coefficients $L_{ij}^{m}$ and $L_{ij}^{c}$ are outputs of KineCluE. $\mathcal{C}$ includes the concentrations of all monomers and clusters:
\begin{equation}
\mathcal{C} = \mathcal{C}_\mathrm{mono} + \mathcal{C}_\mathrm{clusters}  \; .
\label{eq:total_Z}
\end{equation}
The total monomer concentration is given by $\mathcal{C}_\mathrm{mono} = \sum_m \left[m\right] Z_m $, where $\left[m\right]$ is the concentration of monomer $m$ and $Z_m$ its partition function. The total cluster concentration $\mathcal{C}_\mathrm{clusters}$ depends on the specific conditions of the system (e.g., thermal equilibrium or irradiation), and can be obtained for instance with cluster-dynamics models. Under equilibrium conditions and within a dilute approach, the cluster concentrations can be expressed in a low-temperature expansion framework as the product of the concentrations of each cluster component. In this case, $\mathcal{C}_\mathrm{clusters}$ becomes:
\begin{equation}
\mathcal{C}_\mathrm{clusters} = \sum_c \left[c\right] Z_c = \sum_c \left( \prod_k \left[k\right]^{M_k} \right) Z_c \; ,
\end{equation}
where $[k]$ is the concentration of each component $k$ of cluster $c$, and $M_k$ the component multiplicity. For instance, the contribution from a cluster with two vacancies (V) and one solute (B) would be: $\left[\mathrm{V}\right]^2 \left[\mathrm{B}\right] Z_\mathrm{(VVB)}$. The partition functions of each monomer and cluster ($Z_m$ and $Z_c$) are directly output by KineCluE and correspond to the sum of Boltzmann terms $\exp(E_x/k_\mathrm{B}T)$ over all configurations $x$ of the cluster. For monomers, this is equivalent to counting the symmetry-equivalent configurations: for instance, $Z_\mathrm{V}=1$ for vacancies, whereas $Z_\mathrm{I}=6$ for $\langle 110 \rangle$ dumbbells due to their 6 possible orientations in a cubic crystal. The monomer and cluster fractions are then expressed as: 
\begin{align}
f_m & = \frac{\left[m\right] Z_m}{\mathcal{C}} \; ; \label{eq:monomer_fraction_general} \\ 
f_c & = \frac{\left[c\right] Z_c}{\mathcal{C}} \; . \label{eq:pair_fraction_general}
\end{align}

In this study where only one mobile defect monomer $\delta$ (V or I for vacancy and dumbbell, respectively) and one defect-solute pair $\delta \mathrm{B}$ are concerned, the total concentration (Eq. \ref{eq:total_Z}) is given by:
\begin{equation}
\mathcal{C} = \mathcal{C}_\mathrm{mono} + \mathcal{C}_\mathrm{pair} - \mathcal{C}_\mathrm{corr}  \; ,
\label{eq:total_concentration_this_study}
\end{equation}
where the monomer and pair concentrations are respectively $\mathcal{C}_\mathrm{mono} = \left[\delta\right] Z_\delta$ and $\mathcal{C}_\mathrm{pair} = \left[\delta\right] \left[\mathrm{B}\right] Z_{\delta \mathrm{B}}$. $\mathcal{C}_\mathrm{corr}$ is a correction term accounting for the sites that monomers are prevented to occupy due to the geometrical definition of the pair, for kinetic radii well beyond the range of solute-defect thermodynamic interactions. This term amounts to $\left[\delta\right] \left[\mathrm{B}\right] Z_{\delta \mathrm{B}}^0$, where $Z_{\delta \mathrm{B}}^0$ (the non-interacting cluster partition function) corresponds to the count of possible pair geometric configurations within the defined kinetic radius $r_\mathrm{kin}$. With this correction, the total transport coefficients are nearly independent of $r_\mathrm{kin}$. Equation (\ref{eq:total_concentration_this_study}) thus becomes:
\begin{equation}
\mathcal{C} = \left[\delta\right] \left[ Z_\delta + [\mathrm{B}] \left( Z_{\delta \mathrm{B}} - Z_{\delta \mathrm{B}}^0 \right)  \right]  \; ,
\label{eq:total_concentration_factorized}
\end{equation}
where $Z_\delta$, $Z_{\delta \mathrm{B}}$, and $Z_{\delta \mathrm{B}}^0$ are the monomer, pair, and non-interacting pair partition functions computed by KineCluE. The fractions of monomers and pairs are written respectively as:
\begin{align}
f_\delta = \frac{\mathcal{C}_\mathrm{mono}-\mathcal{C}_\mathrm{corr}}{\mathcal{C}} & = \frac{ Z_\delta - \left[\mathrm{B}\right] Z^0_{\delta \mathrm{B}} } {Z_\delta + \left[\mathrm{B}\right] \left( Z_{\delta \mathrm{B}} - Z^0_{\delta \mathrm{B}} \right) }   \; ; \label{eq:monomer_fraction_specific} \\ 
f_{\delta \mathrm{B}} = \frac{\mathcal{C}_\mathrm{pair}}{\mathcal{C}} & = \frac{\left[\mathrm{B}\right] Z_{\delta \mathrm{B}}} {Z_\delta + \left[\mathrm{B}\right] \left( Z_{\delta \mathrm{B}} - Z^0_{\delta \mathrm{B}} \right) } \; . 
\label{eq:pair_fraction_specific}
\end{align}
Thus, the total transport coefficients (Eq. (\ref{eq:total_lij})) are:
\begin{align}
L_{\delta \delta} & = \mathcal{C} \left[ f_\delta L_{\delta\delta}^{(\delta)} + f_{\delta \mathrm{B}} L_{\delta\delta}^{(\delta \mathrm{B})} \right] \; ; \label{eq:total_Ldeltadelta} \\ 
L_{\delta \mathrm{B}} & = \mathcal{C} f_{\delta \mathrm{B}} L_{\delta \mathrm{B}}^{(\delta \mathrm{B})} \; ;  \label{eq:total_LdeltaB} \\ 
L_{\mathrm{B} \mathrm{B}} & = \mathcal{C} f_{\delta \mathrm{B}} L_{\mathrm{B}\mathrm{B}}^{(\delta \mathrm{B})} \; , \label{eq:total_LBB}
\end{align}
where $L_{ij}^{(\delta)}$ and $L_{ij}^{(\delta \mathrm{B})}$ are the monomer and pair cluster transport coefficients yielded by KineCluE. 

The correction term $\mathcal{C}_\mathrm{corr}$ can be applied as long as $f_m>0$. This implies that $C_\mathrm{corr}<C_\mathrm{mono} $, hence yielding a constraint on the maximum solute concentration that can be treated in this dilute model: 
\begin{equation}
\left[\mathrm{B}\right] = c_\mathrm{B} < \frac{Z_\delta}{Z_{\delta\mathrm{B}}^0}  \; .
\end{equation}
This means that the maximum solute concentration is inversely proportional to $Z_{\delta\mathrm{B}}^0$ and thus to the chosen kinetic radius. 

Finally, the solute and defect concentrations depend on the specific target-alloy composition and conditions. They are assumed here to be fixed by the nominal solute concentration ($\left[\mathrm{B}\right]=C_\mathrm{B}$), and by the steady-state defect concentration established under irradiation ($\mathcal{C}=C_\delta^\mathrm{fix}$). The latter is then split into monomers and pairs according to the fractions in Eqs. (\ref{eq:monomer_fraction_specific}) and (\ref{eq:pair_fraction_specific}). This way, the monomer concentration $\left[\delta \right]$ can be derived from $C_\delta^\mathrm{fix}$ via Eq. (\ref{eq:total_concentration_factorized}), and play no role in solute-transport and RIS behaviors, as shown in the formula summary in Table \ref{tab:formula_summary}.  


\begin{table*}[htb]\scriptsize
	\caption{\label{tab:formula_summary} Summary of formulas used to infer transport properties and RIS tendencies from the KineCluE output \cite{nastar_radiation-induced_2012, schuler_transport_2016, martinez_role_2018, schuler_kineclue:_2020}. Subscripts '(V)' and '(I)' refer to monomers (isolated defects), '(VB)' and '(IB)' to solute-defect pairs. $Z_{\delta \mathrm{B}}$ is the pair partition function, while $Z^0_{\delta \mathrm{B}}$ marks the number of possible geometric configurations associated to the $\delta\mathrm{B}$ pair. $Z_\mathrm{V}=1$ and $Z_\mathrm{I}=6$ are the possible geometric configurations of single vacancies and single dumbbells. The thermodynamic factors $\phi$ and $\xi$ are set to 1 and 0, respectively. Total defect concentrations $C_\mathrm{V}^\mathrm{fix}$ and $C_\mathrm{I}^\mathrm{fix}$ are kept as variable parameters. Labels 'V' and 'I' are perfectly interchangeable, except for the host-related coefficients and the partial diffusion coefficient ratio.} 
	\renewcommand{\arraystretch}{1.6}	
	\centering
	\begin{adjustbox}{width=\textwidth}	
		\begin{tabular}{|C{7.8cm}|C{7.8cm}|}
			\hline
			\multicolumn{1}{|c}{\textbf{Dumbbell diffusion}} & \multicolumn{1}{|c|}{\textbf{Vacancy diffusion}} \\
			\hline
			\multicolumn{2}{|c|}{{\footnotesize \textbf{KineCluE output}}} \\
			\hline
			\gape { monomer: $Z_\mathrm{I} = 6$, $ L_\mathrm{II}^\mathrm{(I)}$ } & \gape { monomer: $Z_\mathrm{V}=1 $, $  L_\mathrm{VV}^\mathrm{(V)}$ } \\
			\gape { pair: $  L_\mathrm{II}^\mathrm{(IB)} $, $  L_\mathrm{IB}^\mathrm{(IB)} $, $ L_\mathrm{BB}^\mathrm{(IB)}$, $ Z_\mathrm{IB}$, $Z_\mathrm{IB}^0  $ }   &  \gape { pair: $ L_\mathrm{VV}^\mathrm{(VB)} $, $ L_\mathrm{VB}^\mathrm{(VB)} $, $L_\mathrm{BB}^\mathrm{(VB)}$, $Z_\mathrm{VB}$, $Z_\mathrm{VB}^0 $ } \\
			\hline
			\multicolumn{2}{|c|}{{\footnotesize \textbf{Total transport coefficients }}} \\
			\hline
			$ \mathcal{C} = \left[\mathrm{I}\right] \left[ Z_\mathrm{I} + \left[\mathrm{B}\right] \left( Z_\mathrm{IB} - Z_\mathrm{IB}^0 \right) \right] = C_\mathrm{I}^\mathrm{fix} $ &  $ \mathcal{C} = \left[\mathrm{V}\right] \left[ Z_\mathrm{V} + \left[\mathrm{B}\right] \left( Z_\mathrm{VB} - Z_\mathrm{VB}^0 \right) \right] = C_\mathrm{V}^\mathrm{fix} $ \\
			\addstackgap[4pt] { $ f_\mathrm{I} = \frac{Z_\mathrm{I} - \left[\mathrm{B}\right] Z_\mathrm{IB}^0 }{Z_\mathrm{I} + \left[\mathrm{B}\right] \left( Z_\mathrm{IB} - Z_\mathrm{IB}^0 \right) } $ } & \addstackgap[4pt] { $ f_\mathrm{V} = \frac{Z_\mathrm{V} - \left[\mathrm{B}\right] Z_\mathrm{VB}^0 }{Z_\mathrm{V} + \left[\mathrm{B}\right] \left( Z_\mathrm{VB} - Z_\mathrm{VB}^0 \right) } $ } \\  
			\addstackgap[4pt] { $ f_\mathrm{IB} = \frac{\left[\mathrm{B}\right] Z_\mathrm{IB} }{Z_\mathrm{I} + \left[\mathrm{B}\right] \left( Z_\mathrm{IB} - Z_\mathrm{IB}^0 \right) } $ } & \addstackgap[4pt] { $ f_\mathrm{VB} = \frac{\left[\mathrm{B}\right] Z_\mathrm{VB}}{Z_\mathrm{V} + \left[\mathrm{B}\right] \left( Z_\mathrm{VB} - Z_\mathrm{VB}^0 \right) } $ } \\  		
			$L_\mathrm{BB}^\mathrm{sia} = C_\mathrm{I}^\mathrm{fix} f_\mathrm{IB} L_\mathrm{BB}^\mathrm{(IB)} $ & $L_\mathrm{BB}^\mathrm{vac} = C_\mathrm{V}^\mathrm{fix} f_\mathrm{VB} L_\mathrm{BB}^\mathrm{(VB)} $ \\
			$L_\mathrm{IB} = C_\mathrm{I}^\mathrm{fix} f_\mathrm{IB} L_\mathrm{IB}^\mathrm{(IB)} $ & $L_\mathrm{VB} = C_\mathrm{V}^\mathrm{fix} f_\mathrm{VB} L_\mathrm{VB}^\mathrm{(VB)} $ \\
			\gape { $L_\mathrm{II} = C_\mathrm{I}^\mathrm{fix} \left( f_\mathrm{I} L_\mathrm{II}^\mathrm{(I)} + f_\mathrm{IB} L_\mathrm{II}^\mathrm{(IB)} \right) $ } & \gape { $L_\mathrm{VV} = C_\mathrm{V}^\mathrm{fix} \left( f_\mathrm{V} L_\mathrm{VV}^\mathrm{(V)} + f_\mathrm{VB} L_\mathrm{VV}^\mathrm{(VB)} \right)  $ } \\
			\hline		
			\multicolumn{2}{|c|}{{\footnotesize \textbf{Host ('A')-related transport coefficients}}} \\
			\hline
			\addstackgap[4pt] { $L_\mathrm{AB}^\mathrm{sia} = L_\mathrm{IB} - L_\mathrm{BB}^\mathrm{sia}$ } & \addstackgap[4pt] { $L_\mathrm{AB}^\mathrm{vac} = - L_\mathrm{VB} - L_\mathrm{BB}^\mathrm{vac}$ } \\
			$L_\mathrm{AA}^\mathrm{sia} = L_\mathrm{II} - 2L_\mathrm{AB}^\mathrm{sia} - L_\mathrm{BB}^\mathrm{sia}$ & $L_\mathrm{AA}^\mathrm{vac} = L_\mathrm{VV} - 2L_\mathrm{AB}^\mathrm{vac} - L_\mathrm{BB}^\mathrm{vac}$ \\
			\gape { $L_\mathrm{AI} = L_\mathrm{AA}^\mathrm{sia} + L_\mathrm{AB}^\mathrm{sia} $ } & \gape { $L_\mathrm{AV} = - L_\mathrm{AA}^\mathrm{vac} - L_\mathrm{AB}^\mathrm{vac} $ } \\
			\hline
			\multicolumn{2}{|c|}{{\footnotesize \textbf{Normalized transport coefficients}}} \\
			\hline
			\gape { $\ell_\mathrm{BB}^\mathrm{sia} = \frac{L_\mathrm{BB}^\mathrm{sia}}{C_\mathrm{I}^\mathrm{fix}} \;\;\; ; \;\;\; \ell_\mathrm{IB} = \frac{L_\mathrm{IB}}{C_\mathrm{I}^\mathrm{fix}} \;\;\; ; \;\;\; \ell_\mathrm{II} = \frac{L_\mathrm{II}}{C_\mathrm{I}^\mathrm{fix}} $ } & \gape { $\ell_\mathrm{BB}^\mathrm{vac} = \frac{L_\mathrm{BB}^\mathrm{vac}}{C_\mathrm{V}^\mathrm{fix}} \;\;\; ; \;\;\; \ell_\mathrm{VB} = \frac{L_\mathrm{VB}}{C_\mathrm{V}^\mathrm{fix}} \;\;\; ; \;\;\; \ell_\mathrm{VV} = \frac{L_\mathrm{VV}}{C_\mathrm{V}^\mathrm{fix}} $ } \\
			\gape{ $\ell_\mathrm{AB}^\mathrm{sia} = \frac{L_\mathrm{AB}^\mathrm{sia}}{C_\mathrm{I}^\mathrm{fix}} \;\;\; ; \;\;\; \ell_\mathrm{AI} = \frac{L_\mathrm{AI}}{C_\mathrm{I}^\mathrm{fix}} \;\;\; ; \;\;\; \ell_\mathrm{AA}^\mathrm{sia} = \frac{L_\mathrm{AA}^\mathrm{sia}}{C_\mathrm{I}^\mathrm{fix}} $ } & \gape{ $\ell_\mathrm{AB}^\mathrm{vac} = \frac{L_\mathrm{AB}^\mathrm{vac}}{C_\mathrm{V}^\mathrm{fix}} \;\;\; ; \;\;\; \ell_\mathrm{AV} = \frac{L_\mathrm{AV}}{C_\mathrm{V}^\mathrm{fix}} \;\;\; ; \;\;\; \ell_\mathrm{AA}^\mathrm{vac} = \frac{L_\mathrm{AA}^\mathrm{vac}}{C_\mathrm{V}^\mathrm{fix}} $ } \\
			\hline
			\multicolumn{2}{|c|}{{\footnotesize \textbf{Flux-coupling ratios}}} \\
			\hline
			\addstackgap[4pt] { $ G_\mathrm{I} = \frac{L_\mathrm{IB}}{L_\mathrm{BB}^\mathrm{sia}} = \frac {L_\mathrm{IB}^\mathrm{(IB)}}{L_\mathrm{BB}^\mathrm{(IB)}} $ } ; $ g_\mathrm{I} = \frac{L_\mathrm{IB}^\mathrm{(IB)}}{L_\mathrm{II}^\mathrm{(IB)}}$ & \addstackgap[4pt] { $ G_\mathrm{V} = \frac{L_\mathrm{VB}}{L_\mathrm{BB}^\mathrm{vac}} = \frac {L_\mathrm{VB}^\mathrm{(VB)}}{L_\mathrm{BB}^\mathrm{(VB)}} $ } ; $ g_\mathrm{V} = \frac{L_\mathrm{VB}^\mathrm{(VB)}}{L_\mathrm{VV}^\mathrm{(VB)}}$ \\
			\hline
			\multicolumn{2}{|c|}{{\footnotesize \textbf{Solute tracer diffusion coefficient}}} \\
			\hline
			\addstackgap[4pt] { $ D_\mathrm{B^*}^\mathrm{sia} = \frac{L_\mathrm{BB}^\mathrm{sia}}{C_\mathrm{B}} = C_\mathrm{I}^\mathrm{fix} \left[ \frac{Z_\mathrm{IB} L_\mathrm{BB}^\mathrm{(IB)}}{Z_\mathrm{I} + \left[\mathrm{B}\right] \left( Z_\mathrm{IB} - Z_\mathrm{IB}^0 \right)} \right]$ } & \addstackgap[4pt] { $ D_\mathrm{B^*}^\mathrm{vac} = \frac{L_\mathrm{BB}^\mathrm{vac}}{C_\mathrm{B}} = C_\mathrm{V}^\mathrm{fix} \left[ \frac{Z_\mathrm{VB} L_\mathrm{BB}^\mathrm{(VB)}}{Z_\mathrm{V} + \left[\mathrm{B}\right] \left( Z_\mathrm{VB} - Z_\mathrm{VB}^0 \right)} \right]$ } \\
			\hline		
			\multicolumn{2}{|c|}{{\footnotesize \textbf{Ratio of partial diffusion coefficients}}} \\
			\hline
			\addstackgap[4pt] { $ D_\mathrm{pd}^\mathrm{sia} = \frac{\left(1-C_\mathrm{B}\right)L_\mathrm{BI}}{C_\mathrm{B}L_\mathrm{AI} } = \frac{ \left( 1-\left[ \mathrm{B}\right] \right) f_\mathrm{IB} L_\mathrm{IB}^\mathrm{(IB)}} { \left[ \mathrm{B} \right] \left[ f_\mathrm{I} L_\mathrm{II}^\mathrm{(I)} + f_\mathrm{IB} \left( L_\mathrm{II}^\mathrm{(IB)} - L_\mathrm{IB}^\mathrm{(IB)} \right) \right] }$ } & \addstackgap[4pt] { $ D_\mathrm{pd}^\mathrm{vac} = \frac{\left(1-C_\mathrm{B}\right)L_\mathrm{BV}}{C_\mathrm{B}L_\mathrm{AV} } = - \frac{ \left( 1-\left[ \mathrm{B}\right] \right) f_\mathrm{VB} L_\mathrm{VB}^\mathrm{(VB)}} { \left[ \mathrm{B} \right] \left[  f_\mathrm{V} L_\mathrm{VV}^\mathrm{(V)} + f_\mathrm{VB} \left( L_\mathrm{VV}^\mathrm{(VB)} - L_\mathrm{VB}^\mathrm{(VB)} \right) \right]}$ } \\
			\hline
			\multicolumn{2}{|c|}{{\footnotesize \textbf{Normalized intrinsic diffusion coefficients}}} \\
			\hline
			\addstackgap[4pt] { $  D_\mathrm{A}^\mathrm{sia} = \phi \left(  \frac{\ell_\mathrm{AA}^\mathrm{sia}}{1-C_\mathrm{B}} - \frac{ \ell_\mathrm{AB}^\mathrm{sia} }{C_\mathrm{B} }  -  \frac{ \ell_\mathrm{AI} \xi_\mathrm{AI} } {\phi \left( 1-C_\mathrm{B}\right)} \right) $ } & \addstackgap[4pt] { $  D_\mathrm{A}^\mathrm{vac} = \phi \left(  \frac{\ell_\mathrm{AA}^\mathrm{vac}}{1-C_\mathrm{B}} - \frac{ \ell_\mathrm{AB}^\mathrm{vac} }{C_\mathrm{B} }  -  \frac{ \ell_\mathrm{AV} \xi_\mathrm{AV} } {\phi \left( 1-C_\mathrm{B} \right) } \right) $ } \\
			\addstackgap[4pt] { $  D_\mathrm{B}^\mathrm{sia} =  \phi \left( \frac{\ell_\mathrm{BB}^\mathrm{sia}}{C_\mathrm{B}} - \frac{ \ell_\mathrm{AB}^\mathrm{sia} }{ 1-C_\mathrm{B} }  -  \frac{ \ell_\mathrm{BI} \xi_\mathrm{BI} } {\phi C_\mathrm{B}} \right) $ } & \addstackgap[4pt] { $  D_\mathrm{B}^\mathrm{vac} =  \phi \left( \frac{\ell_\mathrm{BB}^\mathrm{vac}}{C_\mathrm{B}} - \frac{ \ell_\mathrm{AB}^\mathrm{vac} }{ 1-C_\mathrm{B} }  -  \frac{ \ell_\mathrm{BV} \xi_\mathrm{BV} } {\phi C_\mathrm{B}} \right) $ } \\
			\hline
			\multicolumn{2}{|c|}{{\footnotesize \textbf{Radiation induced segregation}}} \\
			\hline
			\multicolumn{2}{|c|}{ \addstackgap[8pt] { $ \nabla C_\mathrm{B} = - \alpha \frac{\nabla C_\mathrm{V}} {C_\mathrm{V}}  \;\;\; , \;\;\;   \alpha  = \frac { \ell_\mathrm{AV} \ell_\mathrm{AI} } {\ell_\mathrm{AI}\left(D_\mathrm{B}^\mathrm{vac} + K D_\mathrm{B}^\mathrm{sia} \right) + \ell_\mathrm{BI} \left( D_\mathrm{A}^\mathrm{vac} + K D_\mathrm{A}^\mathrm{sia} \right) } \left( \frac{\ell_\mathrm{BV}}{\ell_\mathrm{AV}} - \frac{\ell_\mathrm{BI}}{\ell_\mathrm{AI}}  \right)$ } } \\
			\multicolumn{2}{|c|}{$ K = \frac{C_\mathrm{I}}{C_\mathrm{V}} \approx \frac{D_\mathrm{V}}{D_\mathrm{I}} = \frac{\ell_{\mathrm{VV}}}{\ell_{\mathrm{II}}} = \frac{f_\mathrm{V} L_{\mathrm{VV}}^{(\mathrm{V})} + f_{\mathrm{VB}} L_{\mathrm{VV}}^{(\mathrm{VB})}} {f_\mathrm{I} L_{\mathrm{II}}^{(\mathrm{I})} + f_{\mathrm{IB}} L_{\mathrm{II}}^{(\mathrm{IB})}} $ } \\
			\multicolumn{2}{|c|}{} \\
			\hline
			
		\end{tabular}
	\end{adjustbox}
\end{table*}

\section{Multicomponent partition function}
\label{sec:appendix_multicomponent}

The presence of other solutes than species B in a multicomponent alloy can reduce the amount of solute carriers that actively contribute to the transport of B. This can be taken into account in first approximation with a small modification of the framework in \ref{sec:appendix_computation_total_lij}, if the multicomponent effect on the global equilibrium between the vacancy and interstitial population is neglected. To this purpose, the concentration of additional defect-solute clusters '$e$' is added to the total cluster concentration in Eq. \ref{eq:total_Z} as:
\begin{equation}
\mathcal{C} = \mathcal{C}_\mathrm{mono} + \mathcal{C}_\mathrm{clusters}  + \mathcal{C}_\mathrm{ext}  \; ,
\end{equation}
with $\mathcal{C}_\mathrm{ext} = \sum_{e}\left[ e \right] Z_\mathrm{e}$. Consequently, since $\mathcal{C}$ is larger, the fraction of monomers $f_m$ and clusters $f_c$ contributing to the total transport coefficients (Eqs. \ref{eq:monomer_fraction_general} and \ref{eq:pair_fraction_general}) is smaller. In a monomer/pair framework, this writes:
\begin{equation}
\mathcal{C} = \mathcal{C}_\mathrm{mono} + \mathcal{C}_\mathrm{pair} - \mathcal{C}_\mathrm{corr}  + \left( \mathcal{C}_\mathrm{ext} - \mathcal{C}_\mathrm{ext}^\mathrm
{corr} \right) \; ,
\end{equation}
where $\mathcal{C}_\mathrm{ext} = \sum_{\alpha}\left[ \delta \right] \left[ \alpha \right] Z_\mathrm{\delta \alpha}$ and $\mathcal{C}_\mathrm{ext}^\mathrm{corr} = \sum_{\alpha}\left[ \delta \right] \left[ \alpha \right] Z_\mathrm{\delta \alpha}^0$ ($\alpha$ marks all chemical species different from B). Equation \ref{eq:total_concentration_factorized} is now:
\begin{equation}
\mathcal{C} = \left[\delta\right] \left[ Z_\delta + [\mathrm{B}] \left( Z_{\delta \mathrm{B}} - Z_{\delta \mathrm{B}}^0 \right) + \sum_{\alpha \neq \mathrm{B}} \left[ \alpha \right]  \left( Z_{\delta \alpha} - Z^0_{\delta \alpha} \right) \right]  \; .
\end{equation}
The external correction $C_\mathrm{ext}^\mathrm{corr}$ must be then subtracted from the monomer fraction in Eq. \ref{eq:monomer_fraction_specific}: 
\begin{equation}
f_\delta =\frac{ \mathcal{C}_\mathrm{mono} - \mathcal{C}_\mathrm{corr} - \mathcal{C}_\mathrm{ext}^\mathrm{corr}} {\mathcal{C}} \; , 
\end{equation}
while the pair fraction in Eq. \ref{eq:pair_fraction_specific} remains unchanged.

As a practical example, the multicomponent effect on dumbbell-Mn pairs in an Fe-MnNiP alloy can be accounted for by writing $\mathcal{C}$ as:
	\begin{multline}
	\mathcal{C} = \left[\mathrm{I}\right] \Bigl[ Z_\mathrm{I} + [\mathrm{Mn}] \left( Z_{\mathrm{I} \mathrm{Mn}} - Z_{\mathrm{I} \mathrm{Mn}}^0 \right)  + \\ + \left[ \mathrm{Ni} \right]  \left( Z_{\mathrm{I} \mathrm{Ni}} - Z^0_{\mathrm{I} \mathrm{Ni}} \right) + \left[ \mathrm{P} \right]  \left( Z_{\mathrm{I} \mathrm{P}} - Z^0_{\mathrm{I} \mathrm{P}} \right) \Bigr]  \; .
	\end{multline}
	The monomer and dumbbell-Mn pair fractions then become:
	\begin{equation}
	f_\mathrm{I} = \frac{ Z_\mathrm{I} - \left[\mathrm{Mn}\right] Z^0_{\mathrm{I} \mathrm{Mn}} - \left[\mathrm{Ni}\right] Z^0_{\mathrm{I} \mathrm{Ni}} - \left[\mathrm{P}\right] Z^0_{\mathrm{I} \mathrm{P}} } {\mathcal{C} }   \; ; 
	\end{equation}
	\begin{equation}
	f_{\mathrm{I} \mathrm{Mn}} = \frac{\left[\mathrm{Mn}\right] Z_{\mathrm{I} \mathrm{Mn}}} {\mathcal{C} } \; ,
	\end{equation}
	from which the transport coefficients $L_\mathrm{II}$, $L_\mathrm{IMn}$, and $L_\mathrm{MnMn}$ are computed according to Eqs. \ref{eq:total_Ldeltadelta}, \ref{eq:total_LdeltaB}, and \ref{eq:total_LBB}. When one of the external species (e.g., P) has a very strong binding with dumbbells, the corresponding partition function is large ($Z_\mathrm{IP} >> Z_\mathrm{IMn} $). As a consequence, the fraction of dumbbell-Mn pairs $f_\mathrm{IMn}$ becomes small due to the increased total concentration $\mathcal{C}$.

\section{Radiation-induced segregation model}
\label{sec:appendix_ris}

The RIS tendencies are inferred from the transport coefficients with the mean-field model derived by Nastar, Mart\`inez \textit{et al.} \cite{nastar_radiation-induced_2012, martinez_role_2018} from Wiedersich's theory \cite{wiedersich_theory_1979} and used in several following studies \cite{messina_exact_2014, senninger_modeling_2016, piochaud_atomic-based_2016, thuinet_multiscale_2018}. The solute concentration gradient next to defect sinks is related to the vacancy concentration gradient as \cite{senninger_modeling_2016}:
\begin{equation}
\nabla C_\mathrm{B} = - \alpha \frac{\nabla C_\mathrm{V}}{C_\mathrm{V}} \; ,
\end{equation}
where the $\alpha_\mathrm{(RIS)}$ factor is given by:
\begin{gather}
\alpha = \frac{L_\mathrm{AI}L_\mathrm{AV}}{L_\mathrm{AI}D_\mathrm{B} + L_\mathrm{BI} D_\mathrm{A}} \left( \frac{L_\mathrm{BV}}{L_\mathrm{AV}} - \frac{L_\mathrm{BI}}{L_\mathrm{AI}} \right) \; .
\label{eq:alfa_ris_senninger}
\end{gather}
The expressions of the intrinsic diffusion coefficients $D_\mathrm{A}$ and $D_\mathrm{B}$ as functions of the transport coefficients can be found in one of the aforementioned studies \cite{nastar_radiation-induced_2012}. This model represents the steady-state solution of Eq. \ref{eq:flux_equation} where $J_\mathrm{V}=J_\mathrm{I}$, $J_\mathrm{B}=0$, and the chemical potential gradients due to point defects are developed in a low-temperature expansion framework \cite{nastar_radiation-induced_2012}. In addition, the absence of significant sink bias and a low sink density are assumed. Multiple-solute and multiple-defect effects are also neglected. In this dilute-limit approximation, the thermodynamic factor $\phi$ is equal to 1, and the $\xi$ factors can be assumed to be negligible ($\xi \approx 0$). The sign of $\alpha$ determines if solute enrichment (positive) or depletion (negative) occurs, and is fully controlled by the difference of partial diffusion coefficient ratios (last term in parenthesis in Eq. \ref{eq:alfa_ris_senninger}). In the extremely dilute approach ($C_\mathrm{B} \rightarrow 0$), $\alpha$ is roughly proportional to $C_\mathrm{B}$ due to the fact that $L_\mathrm{BV}$ and $L_\mathrm{BI}$ are themselves proportional to $C_\mathrm{B}$; hence, in this case the ratio $\alpha/C_\mathrm{B}$ is nearly constant.

Since each transport coefficient is proportional to the total defect concentrations $\mathcal{C}=C_\delta^\mathrm{fix}$ (Eqs. \ref{eq:total_Ldeltadelta}, \ref{eq:total_LdeltaB}, \ref{eq:total_LBB}), it is convenient to re-arrange Eq. \ref{eq:alfa_ris_senninger} to isolate $C_\mathrm{V}^\mathrm{fix}$ and $C_\mathrm{I}^\mathrm{fix}$ and analyze $\alpha$ in terms of the point-defect concentration ratio. The factor becomes:
\begin{gather}
\alpha = \frac{\ell_\mathrm{AI}\ell_\mathrm{AV}}{\ell_\mathrm{AI}\left( D_\mathrm{B}^\mathrm{vac} + K D_\mathrm{B}^\mathrm{sia} \right) + \ell_\mathrm{BI}\left( D_\mathrm{A}^\mathrm{vac} + K D_\mathrm{A}^\mathrm{sia} \right) } \left( \frac{\ell_\mathrm{BV}}{\ell_\mathrm{AV}} - \frac{\ell_\mathrm{BI}}{\ell_\mathrm{AI}} \right) \; ,
\label{eq:alfa_ris_appendix}
\end{gather}
where each intrinsic diffusion coefficient has been split into a vacancy and interstitial contribution, and normalized by the corresponding defect concentration (cf. Table \ref{tab:formula_summary}). All transport coefficients $\ell_{ij} = L_{ij}/C_\delta^\mathrm{fix}$ are now independent of $C_\mathrm{V}^\mathrm{fix}$, $C_\mathrm{I}^\mathrm{fix}$, and the latter come into play only in the ratio $K=C_\mathrm{I}^\mathrm{fix}/C_\mathrm{V}^\mathrm{fix}$. This allows for a custom choice of the defect concentration model, e.g., the outcome of a time-dependent rate-theory model adapted to the target conditions. In the latter case, the steady-state solution depends on temperature and sink density, and is in most cases equal to $D_\mathrm{V}/D_\mathrm{I}$ \cite{sizmann_effect_1978}. Since the defect diffusion coefficient is given by $D_\delta = L_{\delta\delta}/\mathcal{C}$, and $L_{\delta\delta}$ is proportional to $\mathcal{C}$ (Eq. (\ref{eq:total_Ldeltadelta})), the $K$ factor can be written as:
\begin{equation}
K = \frac{D_\mathrm{V}}{D_\mathrm{I}} = \frac{\ell_{\mathrm{VV}}}{\ell_{\mathrm{II}}} = \frac{f_\mathrm{V} L_{\mathrm{VV}}^{(\mathrm{V})} + f_{\mathrm{VB}} L_{\mathrm{VV}}^{(\mathrm{VB})}} {f_\mathrm{I} L_{\mathrm{II}}^{(\mathrm{I})} + f_{\mathrm{IB}} L_{\mathrm{II}}^{(\mathrm{IB})}} \; ,
\end{equation}
which reduces, for very low solute concentrations ($f_{\delta\mathrm{B}} \rightarrow 0$, $f_\delta \rightarrow 1$), to a simple ratio of monomer transport coefficients:
\begin{equation}
K = \frac{L_{\mathrm{VV}}^{(\mathrm{V})}} {L_{\mathrm{II}}^{(\mathrm{I})}} \; .
\end{equation}

\end{document}